\documentclass[aps,prx,superscriptaddress,longbibliography,twocolumn,floatfix]{revtex4-1}

\usepackage{amsmath}         
\usepackage{amssymb}         
\usepackage{mathtools}       
\usepackage{graphicx}        
\usepackage{hyperref}        
\usepackage[caption=false]{subfig}  

\let\baraccent=\=       

\newcommand{\abs}[1]{\left| #1 \right|}                                   
\newcommand{\avg}[1]{\langle #1 \rangle}                                  
\newcommand{\largeket}[1]{\left| #1 \right>}                              
\newcommand{\ket}[1]{| #1 \rangle}                                        
\newcommand{\matrixel}[3]{\langle #1 | #2 | #3 \rangle}                   
\newcommand{\largeouterp}[2]{\left| #1 \middle> \middle< #2 \right|}      
\newcommand{\outerp}[2]{| #1 \rangle \langle #2 |}                        
\newcommand{\Tr}{\mathrm{Tr}}                                             
\renewcommand{\=}[1]{\stackrel{#1}{=}}                                    

\allowdisplaybreaks[1]

\begin{document}

\title{Lattice model constructions for gapless domain walls between topological phases}

\author{Chenfeng Bao}
\thanks{These two authors contributed equally to this work.}
\affiliation{Perimeter Institute for Theoretical Physics, Waterloo, Ontario, N2L2Y5, Canada}

\author{Shuo Yang}
\thanks{These two authors contributed equally to this work.}
\affiliation{State Key Laboratory of Low-Dimensional Quantum Physics and Department of Physics, Tsinghua University, Beijing 100084, China}
\affiliation{Frontier Science Center for Quantum Information, Beijing, China}

\author{Chenjie Wang}
\affiliation{Department of Physics and HKU-UCAS Joint Institute for Theoretical and Computational Physics, The University of Hong Kong, Pokfulam Road, Hong Kong, China}

 \author{Zheng-Cheng Gu}
 \email{zcgu@phy.cuhk.edu.hk}
   \affiliation{ Department of Physics, The Chinese University of Hong Kong, Shatin, New Territories, Hong Kong}


\begin{abstract}
Domain walls between different topological phases are one of the most interesting phenomena that reveal the non-trivial bulk properties of topological phases. 
Very recently, gapped domain walls between different topological phases have been intensively studied. In this paper, we systematically construct a large class of lattice models for gapless domain walls between twisted and untwisted gauge theories with arbitrary finite group $G$. 
As simple examples, we numerically study 
several finite groups(including both Abelian and non-Abelian finite group such as $S_3$) in $2$D using the state-of-the-art loop optimization of tensor network renormalization algorithm. We also propose a physical mechanism for understanding the gapless nature of these particular domain wall models. 
Finally, by taking advantage of the classification and construction of twisted gauge theories using group cohomology theory, 
we generalize such constructions into arbitrary dimensions, which might provide us a systematical way to understand gapless domain walls and topological quantum phase transitions.
\end{abstract}

\maketitle

\section{Introduction}
Classification and construction topological phases of quantum matter have become an extremely important and interesting subject in modern condensed matter physics. In the past decade, great achievements have been made toward establishing a complete paradigm for understanding topological phases of quantum matter, especially for systems with strong interactions, from the concept of long range entanglement to the classification of topological phases in generic interacting bosonic and fermionic systems.
\cite{chen11a,chen11b,schuch11,bosonlong,wen15,gu2015,fermionlong,kapustin14,freed14,cheng15,Gu2017,davide2017,barkeshli14,lan16}. Nevertheless, our understanding of topological phase transitions is still very limited, especially in higher dimensions.
Until very recently, it has been realized that a certain class of topological phase transitions in $d$ spatial dimensions can be realized as gapless domain walls between topological phases in $d+1$ spatial dimensions\cite{dunghai1,dunghai2}. Such a holographic principle is very attractive since the properties of gapless domain walls are closely related to the bulk properties of topological phases. It is even possible to establish a paradigm towards understanding generic gapless domain walls and topological phase transitions in future. 

It has been known for a long time that domain walls between two topological states usually exhibit extremely intriguing properties.  At very basic level, there are two types of fundamental domain walls: gapped domain walls and gapless domain walls. 
The properties of gapped domain walls can be systematically studied based on the mathematical framework of unitary modular tensor category (UMTC)theory and the corresponding tunneling matrix formulation. The physical nature of gapped domain walls is also well understood in terms of anyon condensation \cite{bais09,kitaev11,levin13,hung14,TLanAnyCon2015}. Nevertheless, the gapless domain walls are much more complicated and harder to understand in general. Hence a systematic way to construct and understand gapless domain walls is very desired.      

In addition, it would be very useful to distinguish two types of gapless domain walls according to their thermal Hall conductance $K_H$: those with $K_H\neq 0$, and those with $K_H=0$. 
The edge modes of various fractional quantum Hall (FQH) states are natural realizations of the first kind of gapless domain walls \footnote{Throughout the whole paper, we will regard vacuum as a trivial topological state. Thus, the boundary of a topological state can be regarded as a special kind of domain wall.}, and they are well understood in terms of chiral conformal field theory(CFT) as well as (perturbative) gravitational anomaly.   
However, the second kind of gapless domain walls with $K_H=0$ are rather unexpected from simple physical considerations.
Very recently, the gapless conditions for $K_H=0$ domain walls among different Abelian FQHs are established in terms of the mathematical concept of Lagrangian subsets\cite{levin13}.
The underlying physical nature of these gapless domain walls can be explained by the so-called global gravitational anomalies.
Moreover, it is widely believed that CFTs will also naturally emerge for the $K_H=0$ gapless domain walls, however, there is still lacking of systematical understanding and concrete lattice model realization for these CFTs.     

In the presence of global symmetries, gaplessness domain walls exist even in the absence of gravitational anomalies(assuming the corresponding global symmetry does not break spontanouesly or explicitly). Most notably, gapless domain walls can be constructed between different symmetry protected topological (SPT) states \cite{hasan10,qi11,bosonlong,lu12,levin12}. For example, gapless domain walls between free fermion topological insulators/superconductors are well understood in terms of massless free Dirac/Majorana fermions. 
Unfortunately, gapless domain walls for interacting SPT states are much harder to construct and only very few special examples are well understood so far\cite{braid,gapless1,gapless2,fermionlong}. 

Another motivation to construct and study gapless domain walls of topological phases is the novel concept of bulk-edge correspondence between topological quantum field theory (TQFT) and conformal field theory (CFT).
The first concrete example is the correspondence between the 3D bulk Chern-Simons theory and the 2D boundary Wess-Zumino-Witten (WZW) model, where the space of quantum states in the bulk TQFT is identified with the space of conformal blocks of the boundary CFT \cite{CS-WZW}.
Such correspondence can be viewed as an implementation of the holographic principle \cite{holography-1,holography-2}.
The discovery of FQH gives rise to much deeper physical understanding of the bulk-edge correspondence,  
where the appearance of boundary exclusion statistics is associated with the corresponding CFT \cite{anyon-cft} which has their origin from anyonic excitations \cite{anyon-fqhe-1,anyon-fqhe-2}
with fractional statistics.   
Mathematically, UMTC also provides a general framework for describing anyons and fractionalized statistics\cite{kitaev}, 
and can also be used to construct TQFT \cite{tqft-tensorcategory}. 
Therefore, a natural setup to further extend the study of bulk-edge correspondence is the gapless domain wall between different topological phases. 
Apparently, domain walls are of particular interest because they are where different anyons in the two bulk topological phases meet. Understanding the dynamics of such ``anyon meetings'' can give us deep insights into 
anyon dynamics as well as topological phase transitions.

In this paper, we systematically construct lattice models of gapless domain walls with $K_H = 0$ 
between twisted and untwisted gauge theories(for arbitrary finite gauge group $G$). Such kind of gapless domain walls are closely related to bulk topological phase transitions  and can be constructed in arbitrary dimensions. 
As a simple example, we illustrate the major steps of constructing such a gapless domain wall between the toric code model and double semion model. Then we study the domain wall model using the state-of-the-art loop-optimization tensor network renormalization (loop-TNR) algorithm \cite{loop-TNR}.
 Surprisingly, we find that the low energy spectrum of the domain wall model is consistent with the $su(2)_1$ Wess-Zumino-Witten model even in the absence of global $SU(2)$ symmetry. We also find a bulk picture to understand the emergence of the $su(2)_1$ CFT on the domain wall. We further study such kind of gapless domain walls between twisted and untwisted gauge models with several other finite group $G$ in $2$D. For Abelian group examples, we find all of them can be described by Luttinger liquid theory with a central charge $c=1$. For non-Abelian group example such as $S_3$, we find the gapless domain wall model can be described by a CFT with central charge $c=2$. We conjecture that the $su(3)_1$ Wess-Zumino-Witten model could be a very good candidate for such a CFT.  

On the other hand, according to the correspondence between twisted gauge theories and SPT models \cite{braid}, such kind of gapless domain walls also naturally arise on the interface between the trivial and non-trivial SPT states, provided that the global symmetry on the domain wall is not broken spontaneously or explicitly. From SPT point of view, the gapless nature of the domain walls is closely related to gauge anomaly which can be systematically classified and constructed via group cohomology theory in arbitrary dimensions.\cite{bosonlong,fermionlong}. Thus, our constructions  
of gapless domain walls between twisted and untwisted gauge theory models can be easily generalized into higher dimensions by using group cohomology theory. We believe that many universal properties of these gapless domain walls could also be classified by group cohomology theory.

The rest of the paper is organized as follows.
In Sec.~\ref{sec.tc_ds}, we start with a simple example - the gapless domain walls between toric code model and double semion model. We find a conformal field theory (CFT) described by the orbifold double $su(2)_1$ Wess-Zumino-Witten model even in the absence of global $SU(2)$ symmetry for such a gapless domain wall. We also study the physical mechanism for the gapless nature of domain wall models. 
In Sec.~\ref{sec.zn-dw}, we review group cohomology and its role in the systematic classification and construction of domain walls between twisted and untwisted gauge theories. We further study examples with several other gauge group.
Finally, there will be a conclusion and a discussion on how to generalize these gapless domain wall models into higher dimensions.

\section{A simple example: gapless domain wall between $\mathbb{Z}_2$ gauge model and twisted $\mathbb{Z}_2$ gauge model}
\label{sec.tc_ds}

\subsection{$\mathbb{Z}_2$ gauge model and twisted $\mathbb{Z}_2$ gauge model}
\label{sec.tc_ds.dual}
\begin{figure}
    \includegraphics[width = 0.4 \textwidth]{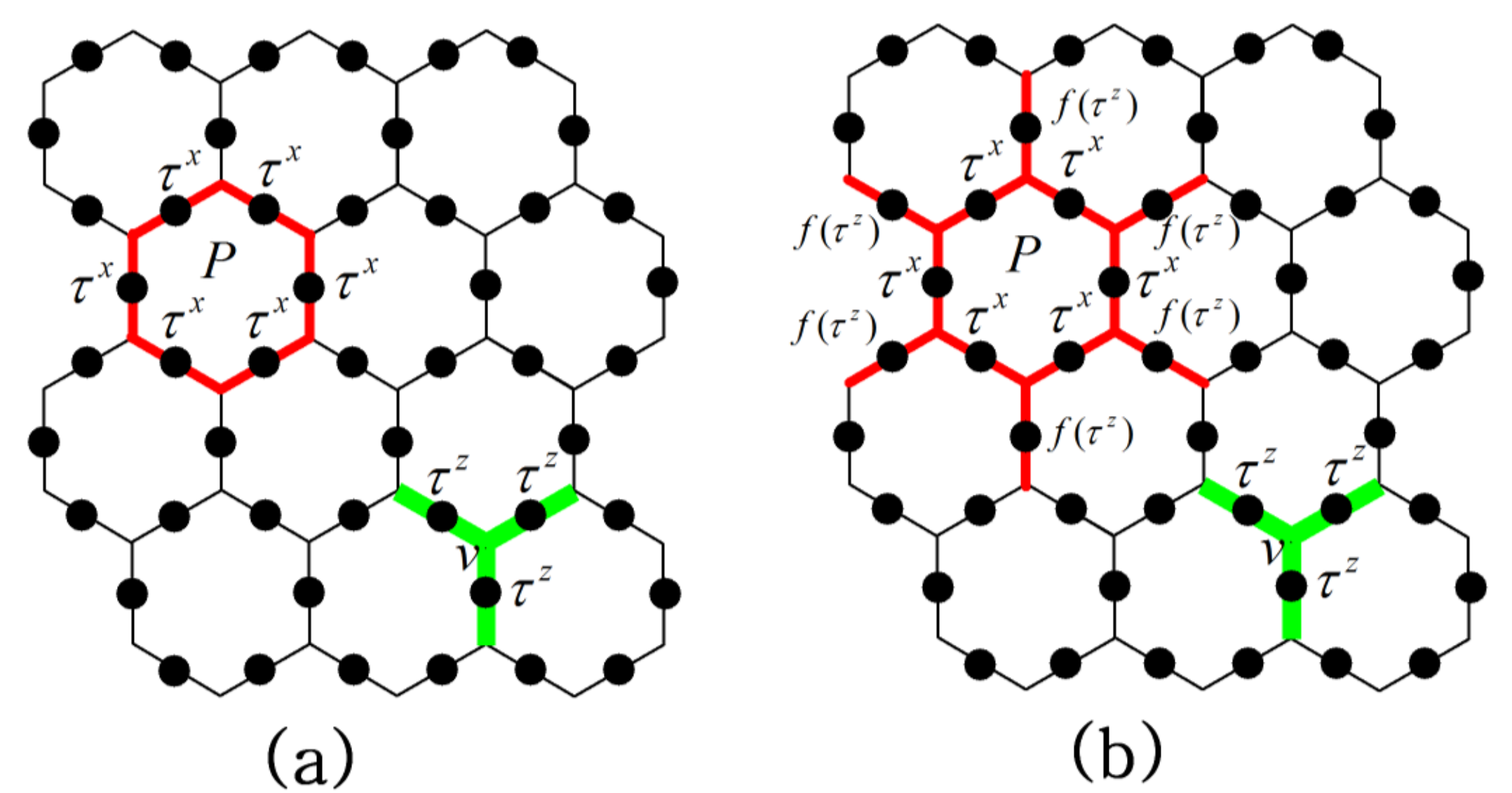}
    \caption{
        The toric code model and doubled semion model $ H_{\mathrm{t.c.}} $, $ H_{\mathrm{d.s.}}$ \eqref{eq.H_tc_ds}.
        For both models, the first term $ Q_v $ is a product of $ \tau_l^z $ on the three links connected to the vertex $ v $,
        the second term is a product of $ \tau_l^x $ on the six links in the plaquette, 
        functions $ f(\tau_l^z) $ on the six links connected to the plaquette,
        and projector $P_p$. 
        (a) $ f(x)=1 $ (trivial) for the toric code model.
        (b) $ f(x)= i^{(1+x)/2} $ for the doubled semion model.
        }
    \label{fig.Hstring}
\end{figure}

Let us begin with the $\mathbb{Z}_2$ quantum double model and twisted quantum double model, namely, the toric code model \cite{TC} and the doubled semion model\cite{DS}.
They can be defined as spin-1/2 systems on a honeycomb lattice where spins live on links. 
The Hamiltonians are (Fig.~\ref{fig.Hstring})
\begin{align} \label{eq.H_tc_ds}
    & H_{\mathrm{t.c.}} = - \sum_v Q_v - \sum_p \left( \prod_{l\in p} \tau_l^x \right) P_p
    \nonumber \\
    & H_{\mathrm{d.s.}} = - \sum_v Q_v - \sum_p \left( \prod_{l\in p} \tau_l^x \prod_{l\in \text{legs of }p} i^{\frac{1+\tau_l^z}{2}} \right) P_p
\end{align}
where \(v, p, l\) denote a vertex, a plaquette, and a link respectively, and
\begin{align}
    Q_v = \prod_{l\in v} \tau_l^z, \qquad  P_p = \prod_{v\in p} \frac{1+Q_v}{2}
\end{align}

Here $\prod_{l\in p}\tau^x_l$ is the product of the $\tau^x_l$ around
a plaquette $p$ and $\prod_{l\in
v}\tau^z_l$ is the product of the $\tau^z_l$ around a vertex $v$.  These two models are simplest examples of string-net models\cite{DS}. 

The ground state $\ket{\Psi_{\mathrm{t.c.}}}$ of
$H_\mathrm{t.c.}$ is exactly known since all the plaquette terms and vertex terms commute with each other.
The string language provides us with a very intuitive way to understand the ground state wave function: we
interpret the $\tau^z_l = -1$ and $\tau^z_l = 1$ states on a single link as
the presence or absence of a string.  (This string is literally an electric
flux line in the $\mathbb{Z}_2$ gauge theory.) The appropriate low energy Hilbert space is
made up of closed string states that satisfy $\prod_{l\in v} \tau^z_l=1$
at every vertex. The ground state is simply a superposition of all closed
string states:
\begin{eqnarray}
\ket{\Psi_{\mathrm{t.c.}}}=\sum_{X \rm{closed}} \ket{X}, \label{Z2wavefunction}
\end{eqnarray}
Such a model realizes the simplest topologically ordered state in 2D.
If we put the ground state wave function Eq.~\eqref{Z2wavefunction} on a torus, there are four different
topological sectors, characterized by even/odd number of large strings wrapping around a torus in both directions.
Moreover, the $\mathbb{Z}_2$ electric charge $e$ can be described as the ends of a string, which are bosons and are created/annihilated in pairs.

$H_\mathrm{d.s.}$ is a less well-known model, which has the same number of ground state degeneracy
on torus, but exhibits a different kind of topological order. The low energy Hilbert space of the model is
again made up of closed string states.  However, the ground state wavefunction
of this model is very similar to the toric code wavefunction except that
different closed string states are weighted by different phase
factors:
\begin{eqnarray}
\label{semWF} \ket{\Psi_{\mathrm{d.s.}}}=\sum_{X \rm{closed}}
(-)^{n(X)}\ket{X}
\end{eqnarray}
where $n$ is the number of closed loops in the closed-string state $X$. The $(-)^{n(X)}$
phase factor makes the $\mathbb{Z}_2$ electric charge(described as the ends of string) carry semion statistics.

\begin{figure}
    \includegraphics[width = 0.4 \textwidth]{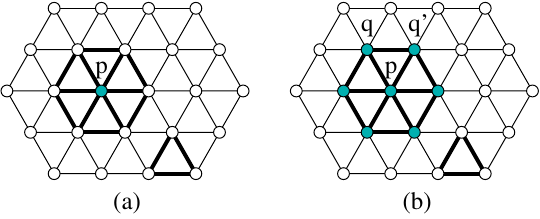}
    \caption{
        $\mathbb{Z}_2$ gauge models $H_0$, $H_1$ \eqref{eq.Hgaugespin}.
        (a) For $H_0$, the first term of the Hamiltonian 
        is a gauge flux term \(\mu_{pq}^z \mu_{qr}^z \mu_{rp}^z \) on the three links of the triangle \( \avg{pqr} \),
        and the second term is a product of $ \sigma_p^x $ and a projector $ O_p $ that acts on the six triangles adjacent to $ p $.
        (b) For $H_1$, the first term is the same gauge flux term, and the second term is more complex $B_p O_p$.
        }
    \label{fig.Hspingauge}
\end{figure}

The above two models can be mapped to $\mathbb{Z}_2$ gauge models and twisted $\mathbb{Z}_2$ gauge models on dual triangular lattices\cite{braid}.
Edges of the triangular lattice are perpendicular to edges of the original honeycomb lattice. Spins on the edges are mapped accordingly. Centers of hexagonal plaquettes in the honeycomb lattice correspond to vertices of the triangular lattice. We put additional spins on these vertices. For each new spin, associate a gauge transformation
\begin{align} \label{eq.gauge-transform}
    W_p = \sigma_p^x \prod_q \mu_{pq}^x
\end{align}
where $p$ labels a vertex in the triangular lattice and $pq$ labels the edge connecting $p$ and $q$, so that dimension of the physical Hilbert space remains the same.
Operator mapping compatible with the gauge transformation is then found to be 
\begin{align}
\tau_l^z = \sigma_p^z \sigma_q^z \mu_{pq}^z, \quad \tau_l^x = \mu_{pq}^x.
\end{align}
The resulting Hamiltonians reads (see Fig.~\ref{fig.Hspingauge})
\begin{gather} \label{eq.Hgaugespin}
\begin{align}
    H_0 = - \sum_{\avg{pqr}} \mu^z_{pq} \mu^z_{qr} \mu^z_{rp} -\sum_p \sigma_p^x O_p \nonumber
    \\
    H_1 = - \sum_{\avg{pqr}} \mu^z_{pq} \mu^z_{qr} \mu^z_{rp} -\sum_p B_p O_p
\end{align}\\
    O_p = \prod_{\avg{pqr}} \frac{1+\mu^z_{pq} \mu^z_{qr} \mu^z_{rp} }{ 2 }
    , \quad
    B_p = \sigma_p^x \prod_{\avg{pqq'}} i^{ \frac{ 1+\sigma_q^z \mu_{qq'}^z \sigma_{q'}^z }{2} } \nonumber
\end{gather}
where the product runs over six triangles adjacent to the vertex $p$.
Apparently, $\mu_{pq}^z$ can be regarded as the $\mathbb{Z}_2$ gauge connection.
Both $H_0$ and $H_1$ are invariant under the $\mathbb{Z}_2$ gauge transformation.

\subsection{Operator algebra for the domain wall between toric code model and double semion model}
\label{sec.tc_ds.H_eff}
\begin{figure}\centering
    \subfloat[]{
        \includegraphics[width = 0.24 \textwidth]{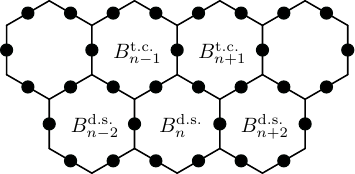}
        \label{fig.hex-interface}
    }
    ~~
    \subfloat[]{
        \includegraphics[width = 0.18 \textwidth]{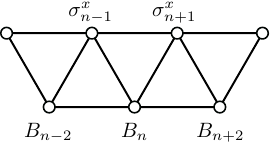}
        \label{fig.gauge-interface}
    }
    \caption{
        (a) Domain wall between the toric code and doubled semion model;
        (b) Effective domain wall between $H_0$ and $H_1$.
    }
\end{figure}

We now consider a system whose upper half plane is described by the toric code model
and lower half plane is described by the doubled semion model.
All local terms of the Hamiltonian commute with each other except on the domain wall.
Denote the plaquette operators in the Hamiltonian Eq.~\eqref{eq.H_tc_ds} as
\begin{align} \label{eq.platq}
    & B^{\mathrm{t.c.}}_p = \prod_{l\in p} \tau_l^x
    \nonumber \\
    & B^{\mathrm{d.s.}}_p = \prod_{l\in p} \tau_l^x \prod_{l\in \text{legs of }p} i^{\frac{1+\tau_l^z}{2}}
\end{align}
and label the plaquettes on the domain wall in a sequential order as shown in Fig.~\ref{fig.hex-interface}.
The nontrivial Hamiltonian algebra on the domain wall can be written as
\begin{align}
    & \left ( B^{\mathrm{t.c.}}_n \right )^{2} = 1, \quad \left ( B^{\mathrm{d.s.}}_n \right )^{2} = 1
    \nonumber \\
    & \left [ B^{\mathrm{t.c.}}_m, B^{\mathrm{d.s.}}_n \right ] = 0, \quad \text{when~} \abs{m-n}>1
    \nonumber \\
    & B^{\mathrm{t.c.}}_n  B^{\mathrm{d.s.}}_{n\pm 1} = 
    - B^{\mathrm{d.s.}}_{n\pm 1} B^{\mathrm{t.c.}}_n \tau_{n-1,n+1}^z \tau_{n\pm 1,n\pm 2}^z
\end{align}
where $ \tau_{m,n}^z $ denotes the spin operator on the edge in between plaquettes $ m $ and $ n $.
Unfortunately, the above Hamiltonian algebra is very complicated and can not be solved in an easy way. 
On the other hand, this domain wall model and its operator algebra can be easily translated to 
the $\mathbb{Z}_2$ gauge models through mappings defined in the previous section (see Fig.~\ref{fig.gauge-interface}):
\begin{align} \label{eq.gauge-int-algebra}
    & \left ( \sigma_n^x \right )^{2} = 1, \quad  ( B_n )^{2} = 1
    \nonumber \\
    & \left [ \sigma_m^x, B_n  \right ] = 0, \quad \text{when~} \abs{m-n}>1
    \nonumber \\
    & \sigma_n^x B_{n\pm 1} = - B_{n\pm 1} \sigma_n^x \sigma_{n\mp 1}^z \sigma_{n\pm 2}^z \mu_{n-1,n+1}^z \mu_{n\pm 1,n\pm 2}^z
\end{align}

\subsection{Gauge fixing and connections to domain wall models between SPT phases}
We would like to simplify the Hamiltonian algebra Eq.~\eqref{eq.gauge-int-algebra} 
in the low energy sector by choosing a convenient gauge and fix all spins on edges.
The $\mathbb{Z}_2$ flux terms  $- \mu^z_{pq} \mu^z_{qr} \mu^z_{rp}$ commute with all other terms in 
both $H_0$ and $H_1$, hence always represent an independent finite energy change.
Therefore, we expect the low energy physics of a gapless domain wall to be entirely captured 
in the subspace where $\mu^z_{pq} \mu^z_{qr} \mu^z_{rp}=1$, i.e. no local $\mathbb{Z}_2$ flux.
We will show later that the domain wall is indeed gapless.
The simplest configuration with no local gauge flux is $\mu^z_{pq} \equiv 1$.
Given any eigenstate of $\{\mu^z_p\}$ with no local gauge flux, the uniform $\mu^z_{pq} \equiv 1$ configuration can be achieved by applying gauge transformations if and only if there is no global gauge flux going through the domain wall, or equivalently in the string-net language, when there is no global string crossing the domain wall.
We will henceforth assume this is the case and use the uniform gauge.

\begin{figure}
    \includegraphics[width = 0.4 \textwidth]{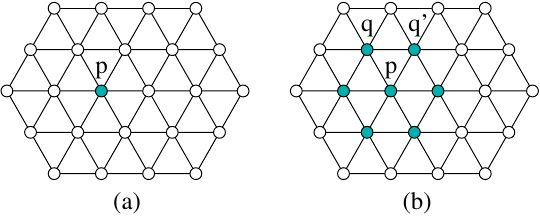}
    \caption{
        Spin models $ H_0, H_1 $ \eqref{eq.H}.
        (a) $ H_0 $ is a sum of all $ \sigma^x_p $.
        (b) $ H_1 $ is a sum of $ \bar B_p = \sigma_p^x \prod_{\avg{pqq'}} i^{\frac{1+\sigma^z_q \sigma^z_{q'}}{2}} $, 
        where the product runs over six triangles $ \avg{pqq'} $ adjacent to site $ p $.
    }
    \label{fig.Hspin}
\end{figure}

Under this uniform gauge, the domain wall Hamiltonian algebra simplifies to
\begin{align}\label{eq.SPT-int-algebra} 
    & \left ( \sigma_n^x \right )^{2} = 1, \quad  ( \bar B_n )^{2} = 1
    \nonumber \\
    & \left [ \sigma_m^x, \bar B_n  \right ] = 0, \quad \text{when~} \abs{m-n}>1
    \nonumber \\
    & \sigma_n^x \bar B_{n\pm 1} = - \bar B_{n\pm 1} \sigma_n^x \sigma_{n\mp 1}^z \sigma_{n\pm 2}^z 
\end{align}
Such a Hamiltonian algebra can naturally arise on the domain wall between the trivial and non-trivial $\mathbb{Z}_2$ SPT phases. The corresponding bulk Hamiltonians are simply those of the gauge models $H_0$, $H_1$ with only spins on vertices and without gauge fields on edges (Fig.~\ref{fig.Hspin})
\begin{gather} \label{eq.H}
\bar{H}_0 = - \sum_p \sigma_p^x, \qquad \bar{H}_1 = - \sum_p \bar{B}_p
\\
\bar{B}_p = \sigma_p^x \prod_{\avg{pqq'}} i^{\frac{1+\sigma^z_q \sigma^z_{q'}}{2}}. \nonumber
\end{gather}
where the product runs over all six triangles $\avg{pqq'}$ containing $p$. 
Both systems have spin-flip $ \mathbb{Z}_2 $ global symmetry $S = \prod_p \sigma_p^x$ inherited from the gauge symmetry, and both have commuting local terms and unique ground states. Specifically, ground state wave functions are
\begin{gather}  \label{eq.gs}
    \ket{\Psi_0} = \sum_{\{\alpha_p\}} \ket{\{\alpha_p\}}   \nonumber \\
    \ket{\Psi_1} = \sum_{\{\alpha_p\}} (-1)^{N_{dw}} \ket{\{\alpha_p\}}
\end{gather}
where $\{\alpha_p\}$ is a spin configuration in $\sigma_p^z$ eigenbasis with $\alpha_p = \uparrow$ or $\downarrow$, 
and $N_{dw}$ is the number of domain walls between spin up and down regions.

As shown in Ref.~\onlinecite{braid}, these two spin models realize the only two short range entangled bosonic phases with onsite $\mathbb{Z}_2$ symmetry. We can, in fact, start from these two SPT models and follow the gauge coupling procedures specified in Ref.~\onlinecite{braid} to obtain the $\mathbb{Z}_2$ gauge models $H_0$, $H_1$.
On the other hand, the domain wall models between the two SPT models $\bar{H}_0$ and $\bar{H}_1$ satisfy the Hamiltonian algebra Eq.~\eqref{eq.SPT-int-algebra} automatically.

\subsection{Effective Hamiltonian with the same operator algebra}
\label{sec.eg-H}
The Hamiltonian algebra of the gauge-fixed flux free domain wall decouples from the bulk terms, so it can be realized on a purely 1+1D spin model. We can straightforwardly check that the following mapping to virtual spin operators $\{\bar{\tau}^x_{n}, \bar{\tau}^y_{n}, \bar{\tau}^z_{n}\}$ preserves this algebra Eq.~\eqref{eq.SPT-int-algebra}
\begin{align}
    \sigma_n^x & = \frac{1}{\sqrt{2}} \left ( \bar \tau^y_n + \bar\tau^z_{n-1}\bar\tau^x_n\bar\tau^z_{n+1} \right )
    \nonumber\\
    \bar B_n & = \frac{1}{\sqrt{2}} \left ( \bar\tau^y_n - \bar\tau^z_{n-1}\bar\tau^x_n\bar\tau^z_{n+1} \right )
    \nonumber\\
    \sigma^z_n & = \bar\tau^z_n
\end{align}
In particular, we verify that
\begin{align}
    & \sigma_n^x \bar B_{n+1}
    \nonumber\\
    = & \frac{1}{2} \left ( \bar\tau^y_{n+1}+\bar\tau^z_{n}\bar\tau^x_{n+1}\bar\tau^z_{n+2} \right ) \left ( \bar\tau^y_n-\bar\tau^z_{n-1}\bar\tau^x_n\bar\tau^z_{n+1} \right )
    \nonumber\\
    = & -\frac{1}{2} \left ( \bar\tau^y_{n+1}-\bar\tau_n^z\bar\tau^x_{n+1}\bar\tau_{n+2}^z \right ) \bar\tau_{n+2}^z\bar\tau^z_{n-1} \left ( \bar\tau^z_{n-1}\bar\tau^x_n\bar\tau^z_{n+1}+\bar\tau^y_n \right )
    \nonumber\\
    = & -\bar\tau^z_{n-1} \bar\tau_{n+2}^z \bar B_{n+1}\sigma_n^x
\end{align}
We further unitarily transform the operators with $U = \prod' \bar\tau^y_n$, where the product runs over every other site on the domain wall. This makes all $\bar B_n$'s and $\sigma_n$'s take the same form and enhances translational symmetry of the effective Hamiltonian, which takes the final form
\begin{align} \label{eq.Hdomainwall}
    H_{\mathrm{dw}} = - \frac{1}{\sqrt{2}} \sum_n 
    \left ( \bar\tau_n^y + \bar\tau_{n-1}^z \bar\tau_n^x \bar\tau_{n+1}^z \right )
\end{align}
We will henceforth refer to this model as the \emph{Ising domain wall} model. This model has a $\mathbb{Z}_2$ symmetry
\begin{align} \label{eq.ising-dw-sym}
S_{\mathrm{dw}} = \prod_n \bar{\tau}^x_n \prod_n \exp\left[\frac{i\pi}{4} \left( \bar{\tau}^z_n \bar{\tau}^z_{n+1} - \bar{\tau}^z_n - 1 \right)\right],
\end{align}
which we will show in Sec.~\ref{sec.zn-dw} is indeed the $\mathbb{Z}_2$ symmetry inherited from the gauge symmetry in the bulk.
This symmetry also acts as a self-dual transformation, as it transforms the two terms $\bar{\tau}^y_n$ and $\bar{\tau}^z_{n-1} \bar{\tau}^x_{n} \bar{\tau}^z_{n+1}$ into each other. We introduce an adjustable parameter $g$ into $H_{\mathrm{dw}}$ and transform this more general model with $S_{\mathrm{dw}}$:
\begin{gather}
H_{\mathrm{dw}}(g)= - \frac{1}{\sqrt{2}}\sum_n \left(g\bar{\tau}^y_n+\bar{\tau}^z_{n-1}\bar{\tau}^x_n\bar{\tau}^z_{n+1} \right)
\\
S_{\mathrm{dw}}^\dag H_{\mathrm{dw}}(g) S_{\mathrm{dw}} = g H_{\mathrm{dw}}(1/g)
\end{gather}
We see that the spectrum of $H_{\mathrm{dw}}(g)$ and $H_{\mathrm{dw}}(1/g)$ is the same up to a factor of $g$, which makes $g=1$ a self-dual point. The spectrum of such a self-dual model is likely to be critical.
On the other hand, since the flux free Hamiltonian algebra Eq.~\eqref{eq.SPT-int-algebra} is also realized as the domain wall between a trivial and a non-trivial $\mathbb{Z}_2$ SPT phases, it is natural to expect such a domain wall model to be gapless if the global $\mathbb{Z}_2$ symmetry is not spontaneously broken.

%

\subsection{Numerical calculations}
\label{sec.domainCFT}

We now perform numerical calculations on the Ising domain wall model $H_{\mathrm{dw}}$ in Eq. \eqref{eq.Hdomainwall}. We show strong evidence that the model is indeed critical. More precisely, our numerical evidence shows that the low-energy physics is described by the $su(2)_1$ WZW theory, or equivalently the compactified free boson CFT at the self-dual radius. 

We perform both exact diagonalization for small system size and loop-TNR calculation for larger system size. For comparison,  we also perform numerical calculations on the spin-1/2 XXX Heisenberg chain
\begin{align}
H_{XXX} = \sum_n \left( \sigma^x_n \sigma^x_{n+1} + \sigma^y_n \sigma^y_{n+1} + \sigma^z_n \sigma^z_{n+1} \right),
\end{align}
which is known to be described by the $su(2)_1$ WZW conformal field theory at low energy.

\begin{figure}
    \includegraphics[width=0.4\textwidth]{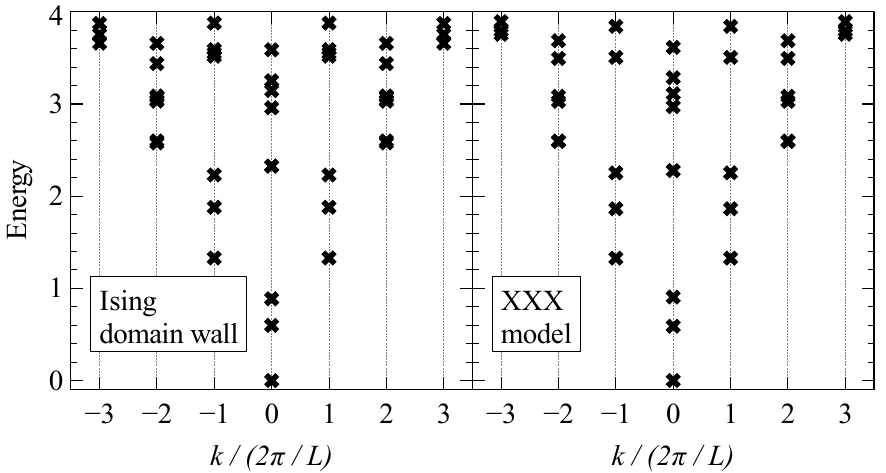}
    \caption{Low energy spectra of Ising domain wall and XXX model at size of 30 sites.}
    \label{fig.spectrum}
\end{figure}

We first compute low energy spectra of both models $H_{\mathrm{dw}}$ and $H_{XXX}$ by exact diagonalization. In Fig.~\ref{fig.spectrum}, the lowest eigenenergies of the two models on a periodic spin chain at the size of 30 sites are plotted against corresponding lattice momenta. 
Both models have a typical CFT excitation tower with linear dispersion, and an identification of low energy states between the two models is clear. Starting from the ground state, degeneracies of the first few energy levels are $\{1, 3, 1, 6, 6, \dots\}$ in both models.
When computing lattice momenta, we use only three-site translations for the Ising domain wall model and two-site translations for the XXX model. Using a finer translational symmetry in either model would cause the unique excitation tower at $k=0$ to split into three/two towers at different momenta, making comparisons difficult.



The $su(2)_1$ WZW theory is equivalent to a free compactified boson at the self-dual radius. A general formula for scaling dimensions of the compactified free boson CFT is
\begin{align}  \label{eq.boson-cft}
\Delta_{m,n} = \frac{1}{2} \left(\rho^2 m^2 + \frac{n^2}{\rho^2} \right), \quad m, n \in \mathbb{Z}.
\end{align}
where $m,n$ label different primary fields and $\rho$ is the compactification radius (in our convention, $\rho=1$ is the self-dual radius). In addition to states with the above scaling dimensions, there are also states corresponding to the current operator as well as its powers and derivatives. The scaling dimensions of the latter states are all integers. More discussion on free compactified boson is given in Appendix \ref{sec:luttinger}. At self-dual radius $\rho=1$, low lying scaling dimensions of all quasi-primaries are $\{0, 0.5, 1.0, 1.5, 2.0, 2.5, \dots\}$ with degeneracies $\{1, 4, 6, 8, 17, 28, \dots\}$. Finite size excitation energies should be proportional to scaling dimensions as $E_i = 2\pi v \Delta_i / L$, but this correspondence can hardly be observed in our exact diagonalization result (Fig.~\ref{fig.spectrum}).
This large deviation from CFT prediction is well understood for the XXX model, where marginally irrelevant fields cause strong finite size effects that fall off only logarithmically with increasing system size\cite{XXXnnn2}. We expect a similar logarithmic convergence for the Ising domain wall.

\begin{figure}
    \includegraphics[width=0.45\textwidth]{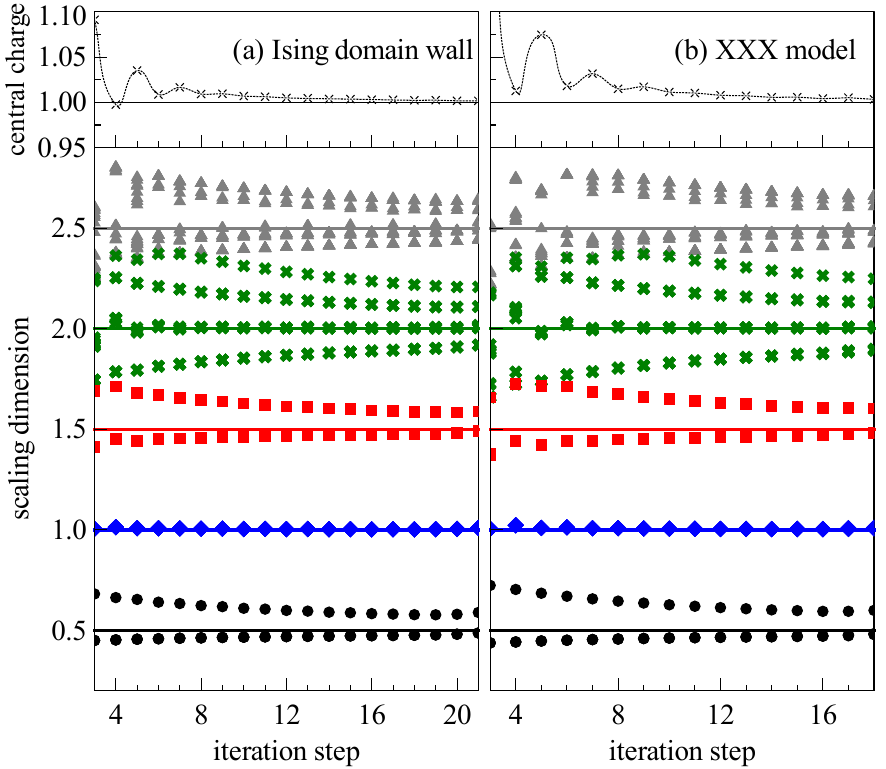}
    \caption{
        Central charge and first 63 nonzero scaling dimensions of Ising domain wall and XXX model computed using loop-TNR. 
        The effective system size grows exponentially with iteration steps. 
        Data points are marked with distinct markers and colors based on their expected converged values.
        Corresponding exact values of \(su(2)_1\) WZW model are shown as solid lines.
        The correct degeneracies of $\{4, 6, 8, 17, 28\}$ are recovered in the expected convergence limit.}
    \label{fig.TNR}
\end{figure}

To access larger system size, we next use loop-TNR\cite{loop-TNR} to compute the central charge and scaling dimensions of both the Ising domain wall and the XXX model. Results are shown in Fig.~\ref{fig.TNR}. The system size grows exponentially with iteration steps in loop-TNR, hence the logarithmic convergence of the finite size effect is translated into a power law, which is indeed observed in Fig.~\ref{fig.TNR} for both models. Even though full convergence cannot be reached before numerical errors drive the system away from the RG fixed point, we can still identify each scaling dimension's corresponding exact values based on their trend, and recover the correct degeneracies in the expected convergence limit.

\begin{figure}
    \includegraphics[width=0.3\textwidth]{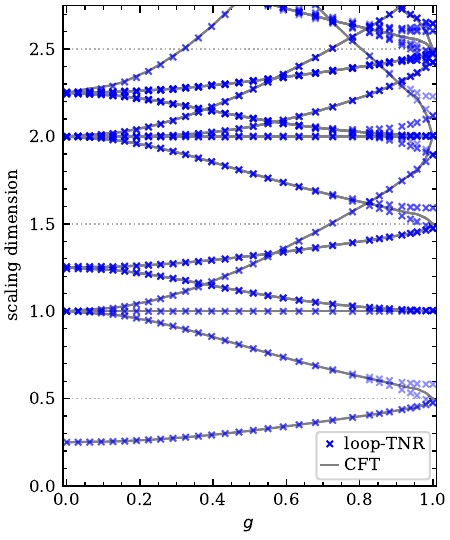}
    \caption{Scaling dimensions of deformed Ising domain wall models \eqref{eq.deform-ising-dw} and corresponding best fit compactified boson CFT predictions \eqref{eq.boson-cft}. Best fit compactification radius $\rho \in [\frac{1}{\sqrt{2}}, 1]$. Agreement is excellent except near $g=1$ where convergence is poor, probably due to large marginally irrelevant operators.}
    \label{fig.boson-fit}
\end{figure}

Finally, we deform the Ising domain wall model without breaking its anomalous global $\mathbb{Z}_2$ symmetry \eqref{eq.ising-dw-sym}, and see how scaling dimensions change with the deformation. 
To find such a suitable deformation, it is easier to first make a unitary transformation
\begin{align}
U = \prod_n \exp\left[ \frac{i\pi}{8} \left( \bar{\tau}^z_n - \bar{\tau}^z_n \bar{\tau}^z_{n+1} \right) \right] \prod_n \bar{\tau}^x_n .
\end{align}
The transformed Hamiltonian can then be written as a special case of a more general class of Hamiltonians all sharing one $\mathbb{Z}_2$ symmetry:
\begin{align}
H'_{\mathrm{dw}}(g) 
= \frac{1}{2} \sum_n \left[ 1 - \bar{\tau}_{n-1}^z \bar{\tau}_{n+1}^z + g \left(\bar{\tau}_{n-1}^z + \bar{\tau}_{n+1}^z\right) \right] \bar{\tau}_n^x  
\label{eq.deform-ising}
\end{align}
with 
\begin{align}
S'_{\mathrm{dw}} = \prod_n \bar{\tau}_n^x \prod_n \exp \left[ \frac{i\pi}{4} \left( \bar{\tau}_n^z \bar{\tau}_{n+1}^z - 1\right)\right]
\label{eq.deform-ising-dw}
\end{align}
In particular, $H'_{\mathrm{dw}}(g=1)$ is the transformed Ising domain wall model, 
and $H'_{\mathrm{dw}}(g=0)$ is unitarily equivalent to the $XY$ model
\begin{gather}
U' = \prod_n \bar{\tau}^x_{4n} \bar{\tau}^x_{4n+1} \prod_n e^{\frac{i\pi}{4}\left(\bar{\tau}^z_{2n}\bar{\tau}^z_{2n+1} - \bar{\tau}^z_{2n}\right)} e^{-\frac{i\pi}{4} \bar{\tau}^x_{2n}} e^{\frac{i\pi}{4}\bar{\tau}^y_{2n+1}}  \nonumber
\\
U'^\dag H'_{\mathrm{dw}}(0) U' = \sum_n \left( \bar{\tau}^x_n \bar{\tau}^x_{n+1} + \bar{\tau}^y_n \bar{\tau}^y_{n+1} \right)
\end{gather}
which is known to realize the $\rho=\frac{1}{\sqrt{2}}$ compactified free boson CFT. We compute scaling dimensions of this deformed Ising domain wall model for $0 \leq g \leq 1$, and find excellent agreement with compactified boson CFT with $\frac{1}{\sqrt{2}} \leq \rho \leq 1$ (Fig.~\ref{fig.boson-fit}).
It is known that for a free boson CFT with $\rho \neq 0.5$, changing $\rho$ is the only relevant direction that preserves both conformal symmetry and $c=1$\cite{ginsparg1988curiosities}. We have therefore rather conclusively shown that the Ising domain wall indeed realizes the compactified free boson CFT at self-dual radius.

\subsection{A physical picture for the gapless nature of the domain wall}
The domain wall between the toric code and double semions can be alternatively viewed as the boundary of a stacking system of the two.
It is then interesting to investigate the bulk properties of the stacking system and how they relate to the boundary domain wall.

We note that the toric code model has four types of anyons $1,e,m,f\equiv em$, where $e,m$ are bosons and the bound state $em$ is a fermion. While the double semion model also has four types of anyons $1,s,\bar{s},b\equiv s\bar{s}$, where $s$ and $\bar{s}$ are semions and the bound state $b$ is a boson. It is well known the toric code model admits a gapped boundary in general since we can condense the Lagrangian subset $(1,e)$ or $(1,m)$. Similarly, the double semion model also admits a gapped boundary since we can condense the Lagrangian subset $(1,b)$. 
Therefore, the stacking systems has sixteen types of anyons described by $(1,e,m,f)\otimes(1,s,\bar{s},b)$. In general, it also admits gapped boundary by  condensing the Lagrangian subsets $(1,e)\otimes(1,b)=(1,e,b,eb)$ or $(1,m)\otimes(1,b)=(1,m,b,mb)$. 

Then what is the mechanism that protects the gapless nature of the domain wall? What makes the domain wall model so special such that it is an $su(2)_1$ WZW CFT? For the first question, a quick answer can be achieved by ungauging the $\mathbb{Z}_2$ gauge symmetry in both models and mapping the domain wall model back to a special boundary of $\mathbb{Z}_2$ SPT phase. Since both toric code model and double semion model can be regarded as the deconfinement phase of $\mathbb{Z}_2$ gauge theory, their domain wall is also in the deconfinement phase where the $\mathbb{Z}_2$ symmetry can not be spontaneously broken. Thus, the corresponding ungauged domain wall model must be a $\mathbb{Z}_2$ SPT boundary without spontaneously symmetry broken, which must be gapless. (It is well known that the boudary of $2+1$D SPT phase must be either gapless or symmetry breaking.) A more rigorous argument can be achieved by regarding the domain wall model as a specific boundary of toric code and double semion stacking systems with $eb$ condensation while $e$ and $b$ are not condensed individually. Clearly, such a condition will exclude the condensation of Lagrangian subset  $(1,e,b,eb)$ or $(1,m,b,mb)$ and protect the gapless nature of the boundary. Very recently, a more precise mathematical language, namely, the categorical symmetry\cite{Wen19} is introduced to understand such a special boundary and our lattice model construction can be regarded as an explicit realization of the maximal categorical symmetry for ground state wevefunctions. 

The second problem is much more subtle, and we need to analyze the anyon content for the stacking model after condensing $eb$. Since all the anyons with nontrivial statistics with $eb$ are confined, the remaining anyons are consisting of one boson $b$ (which is identified to $e$) and two semions $ms, m\bar{s}$ (which are identified with $f\bar{s}, fs$). Together with the identity particle, we end up with a new double semion model with four anyons $1,b,ms,m\bar{s}$, whose corresponding $K$ matrix reads:   
 
\begin{align}
K=\begin{bmatrix}
 2 & 0 \\
 0 & -2
\end{bmatrix}
\end{align}

Apparently, if it is not allowed to condense $b$ in the deconfinement phase, there is no Lagrangian subset can be condensed in the above theory and we will end up with a $c=1$ CFT. It is well known that the above K-matrix describes two layers of filling fraction $\nu=1/2$ bosonic Laughlin states with opposite chirality. Since the edge theory of $\nu=1/2$ bosonic Laughlin state is described by the chiral $su(2)_1$ WZW model, it is quite natural that the domain wall model can be described as the stacking of two chiral $su(2)_1$ WZW models with opposite chirality, if we assume there is no interactions between the left and right movers. However, we stress that the emergence of $SU(2)$ symmetry in the above gapless domain wall model is accidental and it does not apply to generic $\mathbb{Z}_2$ gapless domain wall model preserving the anomalous $\mathbb{Z}_2$ symmetry Eq. (\ref{eq.deform-ising-dw}). Physically, this is because the interactions between left and right movers always exist in generic $\mathbb{Z}_2$ gapless domain wall models, which makes the radius $\rho$ deviate from $1$, e.g., Hamiltonian Eq. (\ref{eq.deform-ising}) with $g \neq 1$. We will discuss more details in next section. 


\section{Constructing lattice models of general gapless domain walls}
\label{sec.zn-dw}
Interesting properties of the Ising domain wall motivate us to construct lattice models of domain walls between more general topological orders, and find their effective field theories.  In the following, we are going to utilize the duality between string-net/gauge models and SPT phases as shown in the $\mathbb{Z}_2$ example, which can be explicitly generalized to arbitrary finite group $G$. 

As with the $\mathbb{Z}_2$ case, a domain wall between  SPT models captures the low energy physics of a corresponding domain wall between gauge models if it is gapless and has no global flux going through.
For simplicity, we will focus on this flux-free case, and directly use the lattice construction of SPT phases to study domain walls between topological phases. We will see that domain walls we construct are all gapless for $G=\mathbb{Z}_N$ case.

\subsection{Constructing SPT phases using group cocycles}
We first briefly review the construction of a lattice model realizing a 2D SPT phase with finite on-site symmetry $G$\cite{bosonlong}.

We define our model on a triangular lattice. Each vertex is associated with a $\abs{G}$-dim Hilbert space where local basis states $\ket{g}$ are labeled by group elements $g\in G$. The model is constructed with a branching structure on the lattice and a $3$-cocycle in the group cohomology $\mathcal{H}^3(G, U(1))$. A branching structure is an assignment of arrows on all edges of the lattice such that there is no local oriented loop, which defines a natural ordering of vertices for each triangle. A $3$-cocycle, for our purpose, is a function $\nu: G^4 \to U(1)$ that satisfies two conditions
\begin{subequations}
\begin{gather}
\nu(gg_0, gg_1, gg_2, gg_3) = \nu(g_0, g_1, g_2, g_3)
\\
\frac{\nu(g_1,g_2,g_3,g_4) \nu(g_0,g_1,g_3,g_4) \nu(g_0,g_1,g_2,g_3)}{\nu(g_0,g_2,g_3,g_4) \nu(g_0,g_1,g_2,g_4)} = 1
\label{eq.cocycle-cdn-simple}
\end{gather}
\end{subequations}
for any $g, g_i \in G$. Two $3$-cocycles $\nu, \nu'$ are considered equivalent if they only differ by a $3$-coboundary $\lambda$, i.e. $\nu' = \nu \lambda$. A $3$-coboundary is a function $\lambda: G^4 \to U(1)$ that satisfies
\begin{gather}
\lambda(g_0,g_1,g_2,g_3) = \frac{\mu(g_1,g_2,g_3) \mu(g_0,g_1,g_3)}{\mu(g_0,g_2,g_3) \mu(g_0,g_1,g_2)} \nonumber
\\
\mu(gg_0,gg_1,gg_2) = \mu(g_0,g_1,g_2)
\end{gather}
The choice of coboundary can be thought of as a gauge freedom for cocycles. Equivalent classes of $3$-cocycles form the third group cohomology $\mathcal{H}^3(G,U(1))$, which itself is an Abelian group. A more detailed introduction to group cohomology can be found in Appendix~\ref{sec.cohomology}.

\begin{figure}
    \includegraphics[width=0.18\textwidth]{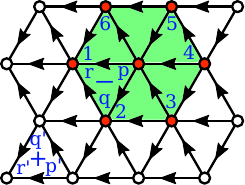}
    \caption{A triangular lattice with a branching structure. The ordering of vertices $pqr$ and the orientation $s_{pqr}=\pm 1$ of a triangle are both in accordance with the branching structure. Each local term of the Hamiltonian acts on seven sites forming a hexagon.}
    \label{fig.spt-triangle-branching}
\end{figure}

Define a unitary transformation on the triangular lattice with a branching structure (Fig.~\ref{fig.spt-triangle-branching})
\begin{align}   \label{eq.spt-defining-unitary}
U_\nu \ket{\{g_i\}} = \prod_{\{pqr\}} \nu^{s_{pqr}}(g_p, g_q, g_r, g^*) \ket{\{g_i\}}
\end{align}
where $g^*$ is a fixed group element, the product runs over all triangles labeled by their three vertices $pqr$ ordered according to the branching structure, and $s_{pqr}=\pm 1$ if the triangle has anticlockwise/clockwise orientation respectively. The Hamiltonian is defined as
\begin{gather}
H_\nu = - \sum_p H_p, \quad H_p = U_\nu \outerp{\phi_p}{\phi_p} U_\nu^\dag \nonumber \\
\ket{\phi_p} = \sum_{g_p \in G} \ket{g}.
\label{eq.H-SPT}
\end{gather}
Although $U_\nu$ acts on the entire lattice, each term of the Hamiltonian acts non-trivially only on seven neighboring sites centered at $p$. Explicitly, with a branching structure as shown in Fig.~\ref{fig.spt-triangle-branching},
\begin{align}
& \matrixel{g_p', g_1 g_2 g_3 g_4 g_5 g_6}{H_p}{g_p, g_1 g_2 g_3 g_4 g_5 g_6} \nonumber \\ 
= & \frac{ \nu(g_4,g_5,g_p,g_p') \nu(g_5,g_p,g_p',g_6) \nu(g_p,g_p',g_6,g_1) }
{ \nu(g_p,g_p',g_2,g_1) \nu(g_3,g_p,g_p',g_2) \nu(g_4,g_3,g_p,g_p') }.
\end{align}
This expression has been simplified using the cocycle condition \eqref{eq.cocycle-cdn-simple}. All local terms commute, so the model is exactly solvable. It has a unique ground state
\begin{align}
\ket{\Psi_{\mathrm{GS}}} = \sum_{\{g_p\}} U_\nu \ket{\{g_p\}}.
\end{align}
Both the Hamiltonian and the ground state have the $G$ symmetry $\{\ket{g_p}\} \to \{\ket{gg_p}\}$.
Models realize distinct SPT phases if and only if they are defined by inequivalent $3$-cocycles.

\subsection{Domain walls between general SPT phases}
We now use the construction in the previous section to derive domain wall models between different SPT phases with the same symmetry $G$.

\begin{figure}
    \includegraphics[width = 0.4\textwidth]{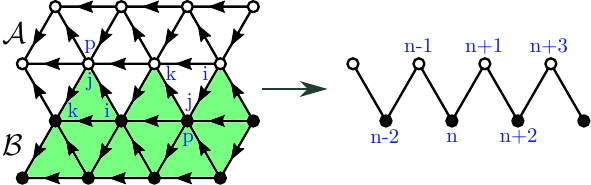}
    \caption{The domain wall of two SPT phases. $H_p^a$ and $H_p^b$ act on white and black vertices respectively. $\mathcal{A}$ denotes the white region, and $\mathcal{B}$ denotes the green shaded region.}
\label{fig.general-dw}
\end{figure}

Consider a system on a triangular lattice with the Hamiltonian
\begin{align}
H = - \sum_{p \in \{\circ\}} H^a_p -\sum_{p \in \{\bullet\}} H^b_p
\end{align}
where $\{\circ\}$ and $\{\bullet\}$ denote vertices in the upper and lower half plane respectively (Fig.~\ref{fig.general-dw}). Each local term $H_p^{a,b}$ acts on seven neighboring sites centered at $p$. $H_p^a$ and $H_p^b$ are defined similarly by Eq.~\eqref{eq.H-SPT}, but use two inequivalent cocycles $\nu_a, \nu_b$.
All $H^a_p$'s commute and all $H^b_p$'s commute, leaving the domain wall as the only nontrivial part of the system.

To explicitly decouple the domain wall from the bulk, we consider the unitary transformation
\begin{align}
U_{ab} = & \prod_{\{pqr\}\in \mathcal{A}} \nu_a^{s_{pqr}}( g_p, g_q, g_r, g^*)  \nonumber \\
     & \cdot \prod_{\{p'q'r'\}\in\mathcal{B}} \nu_b^{s_{p'q'r'}}( g_{p'}, g_{q'}, g_{r'}, g^*)
\end{align}
where $\mathcal{A}$ and $\mathcal{B}$ are regions on the lattice, as shown in Fig.~\ref{fig.general-dw}, that mostly represent upper and lower half planes respectively. 
It changes local terms in the bulk into trivial one body interaction, leaving the domain wall explicitly decoupled. On the domain wall, the Hamiltonian transforms into
\begin{gather}
\label{eq.H-dw-general}
\begin{align}
  U_{ab}^\dag H^a_p U_{ab} & = \sum_{g \in G} \outerp{ g g_p}{g_p} \,
\frac{ \nu^{s_{ijk}}_{ab} (g_i, g_j, g_k, g^*) }{ \nu^{s_{ijk}}_{ab} (g_i, g g_j, g_k, g^*) } , 
\nonumber\\
  U_{ab}^\dag H^b_p U_{ab} & = \sum_{g \in G} \outerp{ g g_p}{g_p} \, 
\frac{ \nu^{s_{ijk}}_{ab} (g_i, g g_j, g_k, g^*) }{ \nu^{s_{ijk}}_{ab} (g_i, g_j, g_k, g^*) } , 
\end{align}
\\
\nu_{ab} \equiv \nu_a^{-1} \nu_b . \nonumber
\end{gather}
$i,j,k$ label a triangle whose top or bottom vertex is $p$, while which vertex each of them represents is assigned according to the branching structure (Fig.~\ref{fig.general-dw}). Due to the group structure of group cohomology $\mathcal{H}^3(G,U(1))$, we see that the domain wall is defined only by one $3$-cocycle $\nu_{ab}$ rather than two.

In general, different branching structures lead to different domain wall models, and $H_p^{a,b}$'s can be different from one another. For concreteness, we will mostly focus on the specific branching structure shown in Fig.~\ref{fig.general-dw}. This branching structure is particularly nice, since it gives us a fully translational invariant domain wall. The corresponding Hamiltonian reads
\begin{gather}
    H = - \sum_n H_n ,  \nonumber \\
    \begin{align}
\label{eq.H-dw-spec-branch}
    & \matrixel{g'_n, g_{n-1} g_{n+1}}{H_n}{g_n, g_{n-1} g_{n+1}} \nonumber \\
    = & \frac{ \nu_{ab}(g_{n+1}, g'_n, g_{n-1}, g^*) }{ \nu_{ab}(g_{n+1}, g_n, g_{n-1}, g^*) } .
    \end{align}
\end{gather}
Its effective anomalous symmetry operator can also be computed
\begin{align} \label{eq.sym-general}
S_g \ket{ \{ g_n \} } = \prod_n \nu_{ab}^{-1} (g_{n+1}, g_n, g^{-1}g^*, g^*) \ket{ \{ g g_n \} }.
\end{align}

Now we have obtained a general domain wall model of 2D bosonic SPT phases. To construct an explicit model, we only need to find explicit expression of group $3$-cocycles.

\subsection{$\mathbb{Z}_N$ domain wall models}
It is no coincidence that both $\mathbb{Z}_N$ gauge theories and SPT phases in 2+1D are classified by the group cohomology $\mathcal{H}^3(\mathbb{Z}_N, U(1)) \cong \mathbb{Z}_N$. We note that string-net realizations of all $N$ distinct $\mathbb{Z}_N$ gauge theories can be constructed\cite{lin2014general-string-net}, and lattice models exist for any SPT phases with finite on-site symmetry\cite{bosonlong}.

The formula of $\mathbb{Z}_N$ $3$-cocycles is well-known\cite{moore1989classical}
\begin{gather}  \label{eq.cocycle-formula}
\nu_3(g_0, g_1, g_2, g_3) 
= \exp\left[i\frac{2\pi k}{N^2} g_{10} \left(g_{21} + g_{32} - g_{31}\right)\right]
\\
g_{ij} \equiv (g_i - g_j)~\mathrm{mod}~N  \nonumber
\\
g_i, k \in \{0,1,\dots,N-1\}  \nonumber
\end{gather}
where $g_i$ labels a group element and $k$ labels the $N$ different classes of $3$-cocycles in $\mathcal{H}^3(\mathbb{Z}_N, U(1)) \cong \mathbb{Z}_N$.
Using this formula in Eq.~\eqref{eq.H-dw-spec-branch}, we can define a $\mathbb{Z}_N$ domain wall model that is labeled by $(N,k)$. Hamiltonians given by $(N,k)$ and $(N,N-k)$ are related by complex conjugation, therefore there are effectively only $\lfloor N/2 \rfloor$ distinct nontrivial domain wall models for a given $N$.

\subsubsection{Re-deriving the $\mathbb{Z}_2$ domain wall}
The simplest nontrivial domain wall model defined by \eqref{eq.H-dw-spec-branch} is given by $(N,k)=(2,1)$. We expect this model to be equivalent to the Ising domain wall we studied in Sec.~\ref{sec.tc_ds}. The cocycle formula in this case simplifies to
\begin{align}
\nu(\{g_i\})
= \begin{dcases}
-1, & \{g_i\} = \{0,1,0,1\} \text{~or~} \{1,0,1,0\} \\
1,  & \text{otherwise}
\end{dcases}
\end{align}
Substituting this formula into Eq.~\eqref{eq.H-dw-spec-branch}, we find
\begin{gather}
H_n = \frac{1}{2}(1 + \sigma^z_{n-1} + \sigma^z_{n+1} - \sigma^z_{n-1} \sigma^z_{n+1} ) \sigma^x_{n},
    \label{eq.H-trans-n1}  \nonumber
\\
S = \prod_n \sigma^x_{n} \prod_n \exp\left[ \frac{i \pi}{4} \left(\sigma^z_{n} \sigma^z_{n+1} - 1 \right)\right], 
\end{gather}
which is exactly the same as Eq.~\eqref{eq.deform-ising-dw} at $g=1$, hence equivalent to the Ising domain wall model.

\subsubsection{$\mathbb{Z}_3, \mathbb{Z}_4$ and $\mathbb{Z}_5$}
We numerically investigate all five distinct domain wall models as defined by \eqref{eq.H-dw-spec-branch} for $N = 3,4,5$. 

\begin{figure}
    \includegraphics[width=0.3 \textheight]{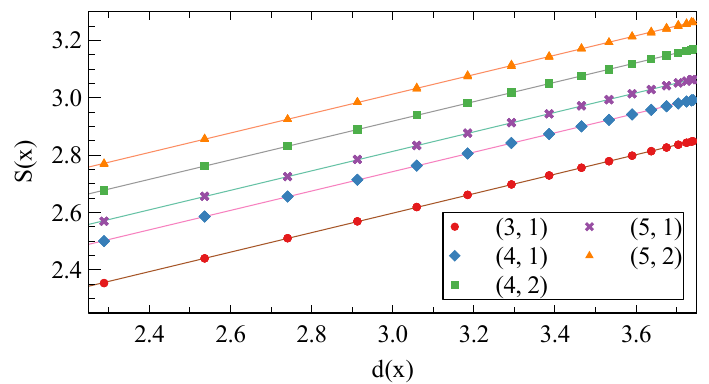}
    \caption{Entanglement entropy $S(x)$ of an interval of length $x$ in the ground state of $L=48$ periodic $\mathbb{Z}_N$ domain wall models labeled by $(N,k)$, plotted against $d(x) = \log(\frac{L}{2\pi}\sin(\frac{x}{L}2\pi))$. Data is fitted with $S(x) = \frac{c}{3} d(x) + b$ and find $c=1.014\textup{--}1.019$ for all models.}
    \label{fig.entropy-zn-domainwalls}
\end{figure}

Entanglement entropy scaling\cite{calabrese2004entanglement} of a 48-site periodic chain as computed by density matrix renormalization group (DMRG)\cite{white1992density,white1993density,verstraete2004density} is plotted in Fig.~\ref{fig.entropy-zn-domainwalls}. We find very precise logarithmic scaling that is fitted with central charge very close to $1$ for all models, proving criticality of these models.

\begin{figure}
    \includegraphics[width = 0.48 \textwidth]{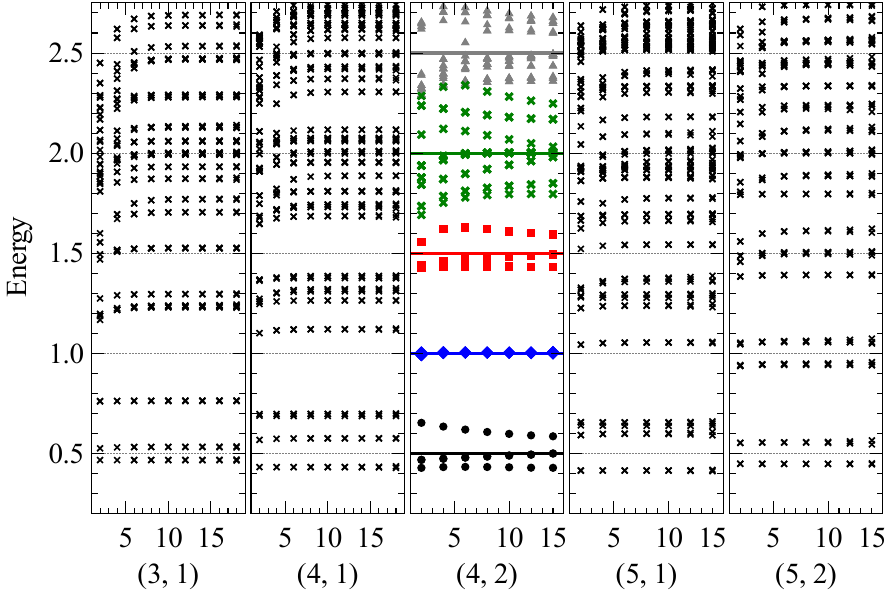}
    \caption{Low lying excitation energies of $\mathbb{Z}_N$ domain wall models labeled by $(N,k)$, plotted against iteration step, which is logarithmic scale in system size. Data points for the $(4,2)$ model are marked with different shapes and colors according to their expected convergence limits, whose values are marked with solid horizontal lines.}
    \label{fig.tnr-zn-domainwalls}
\end{figure}

We next use loop-TNR to compute lowest virtual energies of these models, normalized such that ground state energy is $-1/12$, compatible with a $c=1$ CFT. See Fig.~\ref{fig.tnr-zn-domainwalls}.
Here, we define \emph{virtual energies} via the virtual-space transfer matrix\cite{suzuki1985transfer} of the quantum lattice model. They are characterized by a theory relating to the original lattice model by an $\mathcal{S}$ modular transformation. See Appendix~\ref{sec.comp-details} for details.

The loop-TNR computation does not converge for the $(4,2)$ model, but its converging behavior is very similar to that of the Ising domain wall model. We again see converging trends towards $\{0.5, 1.0, 1.5, 2.0, 2.5\}$ and degeneracies $\{4,6,8,17,28\}$ in the expected limit. We hence conjecture that it is also described by the compactified free boson CFT at self-dual radius.

The other four models' virtual energies do not match any known $c=1$ CFT\cite{ginsparg1988curiosities} (in the usual Euclidean space with the metric being the identity matrix). The gapless edge of an Abelian topological phase is expected to be described by a Luttinger liquid action\cite{wen1995topological}
\begin{align} \label{eq.LL}
S_{\mathrm{edge}} = \frac{1}{4\pi} \int dt dx \left( K_{IJ} \partial_t \phi_I \partial_x \phi_J - V_{IJ} \partial_x \phi_I \partial_x \phi_J \right)
\end{align}
where $K$ is an integer symmetric matrix, $V$ is a positive definite symmetric matrix, and fields are compactified $\phi_I = \phi_I + 2\pi$. Our domain wall models can be viewed effectively as special edge models. Since $c=1$ from the entanglement entropy scaling, we expect $K$ to be a $2\times 2$ matrix with eigenvalues of opposite signs. In addition, since the domain wall models are effectively 1D lattice models, we required the edge theory to be modular invariant.  Hence, we are led to the conclusion that by redefining the fields $\phi_I$ through a linear combination, we can simultaneously diagonalize both matrices with a congruent transformation such that
\begin{align}
PKP^T = \begin{pmatrix} 1 & 0 \\ 0 & -1 \end{pmatrix}, \quad
PVP^T = \begin{pmatrix} v_1 & 0 \\ 0 & v_2 \end{pmatrix}.
\end{align}
This diagonalized form is equivalent to two compactified massless free chiral bosons moving in opposite directions with velocities $v_1$ and $v_2$ respectively. In the usual CFT defined with the Euclidean metric, conformal invariance requires $v_1=v_2$. In our case, $v_1=v_2$  is in general not satisfied. For the general $v_1\neq v_2$ case, we find that energies of highest weight states are (see Appendix \ref{sec:luttinger})
\begin{subequations} \label{eq.chiral-boson-energies}
\begin{align}
E_{m,n} = \frac{1}{2} \left(\rho^2 m^2 + \frac{n^2}{\rho^2} \right) + \frac{1-v_r}{1+v_r} mn ,
\end{align}
and energies of descendant states are
\begin{align}
E_{m,n,\{n_{1,l}\},\{n_{2,l}\}} = E_{m,n} + \frac{2}{1+v_r} \sum_{l=1}^\infty l \left( n_{1,l} + v_r n_{2,l} \right)
\end{align}
\end{subequations}
in units of $\pi(v_1+v_2)/L$, where $v_r = v_1/v_2$. We note that an $\mathcal{S}$ modular transformation has the effect of mapping $v_r \to 1/v_r$, which does not change the spectrum. Hence if virtual energies of a lattice model are characterized by Luttinger liquid, the lattice model itself should be described by the same theory.
We proceed to fit the energies in Fig.~\ref{fig.tnr-zn-domainwalls} with Eq.~\eqref{eq.chiral-boson-energies} by adjusting the two free parameters $\rho$ and $v_r$, and find a perfect fit for all models. The fitted values are listed in Table~\ref{tab.chiral-fit-tnr}. 

\begin{table}
    \begin{tabular}{c|@{\hspace{.3cm}}c @{\hspace{.3cm}} c}
        $(N,k)$ & $\rho$ & $v_1/v_2$ \\
        \hline
        $(3,1)$ & 0.968 & 1.623 \\
        $(4,1)$ & 0.931 & 1.907 \\
        $(5,1)$ & 0.912 & 2.130 \\
        $(5,2)$ & 0.948 & 1.121
    \end{tabular}
    \caption{Best fit parameters for non-conformal $\mathbb{Z}_N$ domain wall models as described by a chiral boson theory \eqref{eq.chiral-boson-energies}.}
    \label{tab.chiral-fit-tnr}
\end{table}

\begin{figure}
    \includegraphics[width=0.4\textwidth]{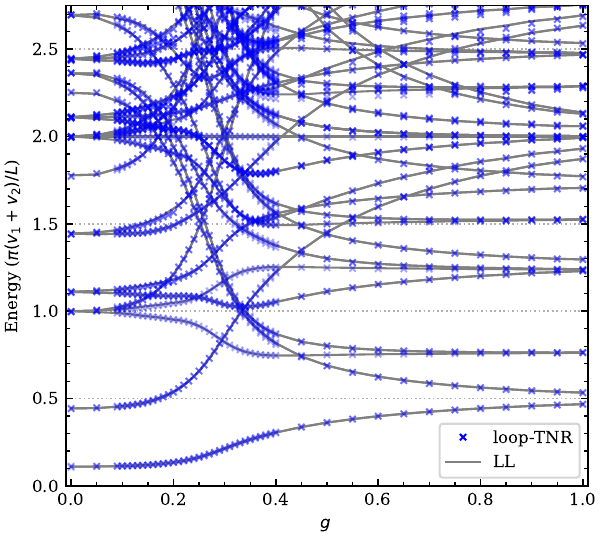}
    \caption{Excitation energies of $H_{\mathrm{deform}}(g)$ and corresponding best fit Luttinger liquid (LL). States with the same quantum numbers $m,n$ in \eqref{eq.chiral-boson-energies} are connected by gray lines.}
    \label{fig.tnr-z3-deform}
\end{figure}

To further confirm this field theoretic description, we continuously deform the $(3,1)$ domain wall model without breaking the effective $\mathbb{Z}_3$ symmetry defined by \eqref{eq.sym-general}, and connect it with a compactified free boson CFT for which $v_1=v_2$. One such deformation is found to be
\begin{gather}
H_{\mathrm{deform}}(g) = g \left(H_{(3,1)} + I \right) + (1-g) H'
\end{gather}
with
\begin{gather}
H' = - \sum_i H'_i \nonumber \\
\begin{align}
H'_i = & \largeouterp{020}{010} + e^{i\frac{2\pi}{3}} \largeouterp{101}{121} \nonumber \\
& + e^{-i \frac{2\pi}{3}} \largeouterp{212}{202} + \mathrm{h.c.}
\end{align}
\end{gather}
where $H_{(3,1)}$ is the domain wall Hamiltonian, and $H'$ is a fine tuned Hamiltonian that realizes the $\rho=\sqrt{2}/3$ compactified free boson CFT.
This deformed model realizes a CFT at $g=0$, and recovers the $(3,1)$ domain wall model (up to a constant) at $g=1$.
Virtual energies of $H_{\mathrm{deform}}(g)$ are computed with loop-TNR for $0 \leq g \leq 1$ (Fig.~\ref{fig.tnr-z3-deform}) and fitted with the field theoretic predictions Eq.~\eqref{eq.chiral-boson-energies} to excellent agreement.

\subsubsection{Alternative branching structure and cocycle gauges}
Models we studied in the previous section all assumed a particular branching structure (Fig.~\ref{fig.general-dw}) and cocycle gauge \eqref{eq.cocycle-formula}. Changing these choices amounts to a local unitary transformation in the bulk. This is inconsequential for the bulk physics, but can cause nontrivial changes on the domain wall, because the resulting local unitary transformations on the two sides may not be the same.
It is important that our general conclusions remain valid for different choices of branching structures and cocycle gauges.

\begin{figure}
    \includegraphics[width=0.18\textwidth]{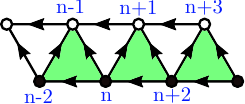}
    \caption{An alternative branching structure on the domain wall.}
    \label{fig.alt-branching}
\end{figure}

Consider an alternative branching structure on the domain wall, shown in Fig.~\ref{fig.alt-branching}. Applying this branching structure to Eq.~\eqref{eq.H-dw-general}, the new domain wall Hamiltonian for a general group cocycle reads
\begin{gather}
H' = - \sum_n \left(H^a_{2n} + H^b_{2n+1}\right) ,  \nonumber \\
\begin{align}
& \matrixel{g'_n, g_{n-1} g_{n+1}}{H^a_n}{g_n, g_{n-1} g_{n+1}} \nonumber \\
= & \frac{ \nu_{ab}(g_{n+1}, g_{n-1}, g_n, g^*) }{ \nu_{ab}(g_{n+1}, g_{n-1}, g'_n, g^*) } ,
\\
& \matrixel{g'_n, g_{n-1} g_{n+1}}{H^b_n}{g_n, g_{n-1} g_{n+1}} \nonumber \\
= & \frac{ \nu_{ab}(g_n, g_{n+1}, g_{n-1}, g^*) }{ \nu_{ab}(g'_n, g_{n+1}, g_{n-1}, g^*) } .
\end{align}
\label{eq.H-dw-alt-branch}
\end{gather}

Using the same cocycle gauge \eqref{eq.cocycle-formula}, the new $\mathbb{Z}_2$ domain wall Hamiltonian is
\begin{align}  \label{eq.H-z2-alt}
H^a_n &= \frac{1}{2} \left(1-\sigma^z_{n-1}+\sigma^z_{n+1}+\sigma^z_{n-1}\sigma^z_{n+1}\right)\sigma^x_n , \nonumber \\
H^b_n &= \frac{1}{2} \left(1+\sigma^z_{n-1}-\sigma^z_{n+1}+\sigma^z_{n-1}\sigma^z_{n+1}\right)\sigma^x_n.
\end{align}
This is unitarily equivalent to the Ising domain wall. Explicitly, Eq.~\eqref{eq.H-z2-alt} is related to Eq.~\eqref{eq.deform-ising-dw} at $g=1$ by
\begin{align}
U'' = \prod_n \exp\left( \frac{i\pi}{4} \sigma^z_{2n}\sigma^z_{2n+1} \right) \exp\left( -\frac{i\pi}{4} \sigma^z_n \right).
\end{align}

The $\mathbb{Z}_3$ domain wall \emph{is} changed by this new branching structure. We again compute its virtual energies, see Fig.~\ref{fig.alt-branching}, and find that it still fits the Luttinger liquid predictions perfectly at $\rho=0.951, v_r=1.416$.

\begin{figure}
    \includegraphics[width=0.2\textwidth]{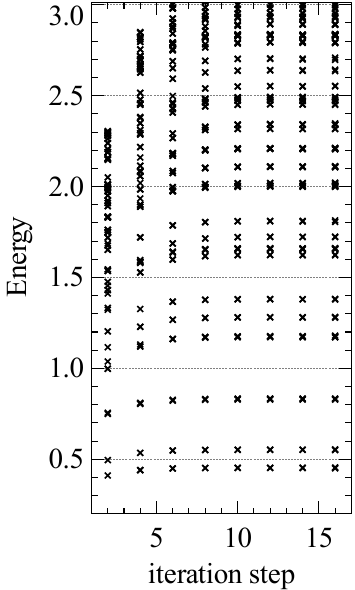}
    \caption{Excitation energies of the $\mathbb{Z}_3$ domain wall model with alternative branching structure (Fig.~\ref{fig.alt-branching}).}
\end{figure}

\begin{figure}
    \includegraphics[width=0.48\textwidth]{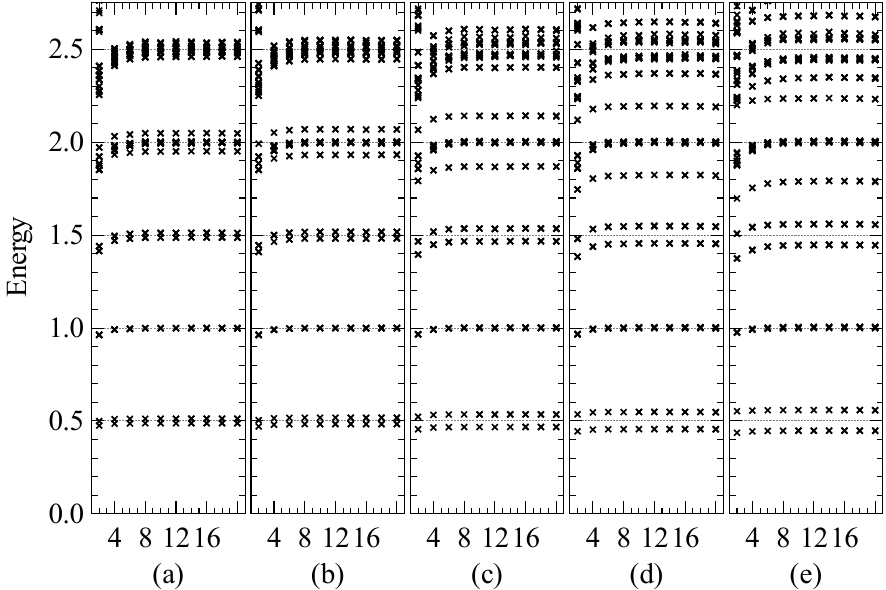}
    \caption{Excitation energies/scaling dimensions of $\mathbb{Z}_2$ domain walls defined with random $3$-coboundaries, plotted against loop-TNR iteration number. All models match free boson CFT with compactification radius $\rho \in [0.945, 1.0]$.}
    \label{fig.tnr-z2-coboundary}
\end{figure}

\begin{figure}
    \includegraphics[width=0.48\textwidth]{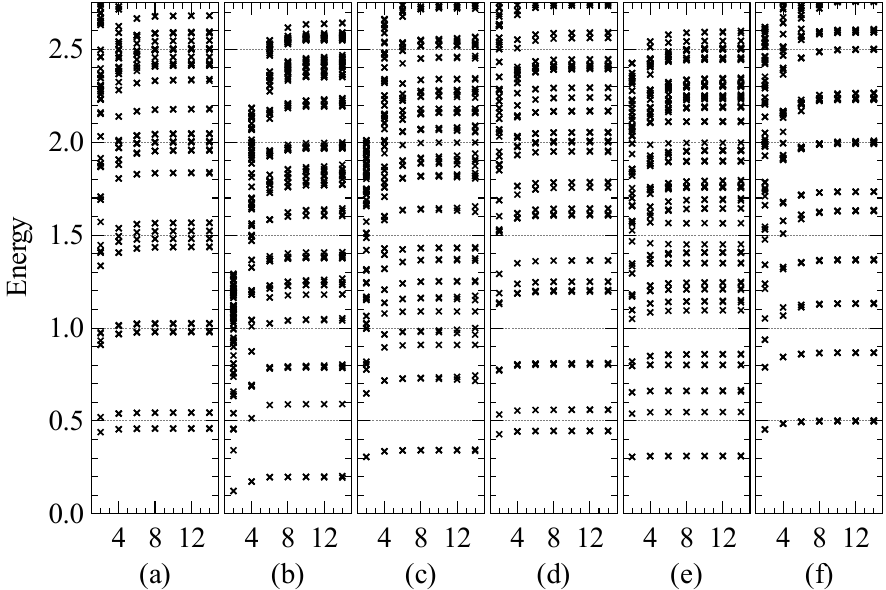}
    \caption{Excitation energy of $\mathbb{Z}_3$ domain walls defined with random $3$-coboundaries, plotted against loop-TNR iteration number. All computed models match a Luttinger liquid/chiral boson theory with appropriate parameters (Table~\ref{tab.z3-coboundary-fit}).}
    \label{fig.tnr-z3-coboundary}
\end{figure}

Next, we keep the old branching structure in Fig.~\ref{fig.general-dw}, but multiply each $3$-cocycle in \eqref{eq.H-dw-spec-branch} with an arbitrary $3$-coboundary, effectively changing the cocycle gauge. Any $\mathbb{Z}_N$ group coboundary is fully determined by $N^2$ independent parameters:
\begin{align}
\mu(0,m,n) = e^{i\theta_{mn}}
\end{align}
For both $\mathbb{Z}_2$ and $\mathbb{Z}_3$, we generate 16 sets of random $\theta_{mn} \in [0,2\pi)$, and compute virtual energies of resulting domain wall models with loop-TNR. All generated $\mathbb{Z}_2$ domain walls fit a free boson CFT with compactification radius $\rho \in [0.945, 1]$, and all generated $\mathbb{Z}_3$ domain walls fit a Luttinger liquid theory with appropriate $\rho$ and $v_1/v_2$. Energies of a sample of these models are shown in Fig.~\ref{fig.tnr-z2-coboundary} and Fig.~\ref{fig.tnr-z3-coboundary}. Best fit parameters for $\mathbb{Z}_3$ domain walls are listed in Table~\ref{tab.z3-coboundary-fit}.

\begin{table}
    \begin{tabular}{c|@{\hspace{.3cm}}c @{\hspace{.3cm}} c}
         & $\rho$ & $v_1/v_2$ \\
        \hline
        (a) & 0.958 & 1.045 \\
        (b) & 0.631 & 2.391 \\
        (c) & 0.827 & 1.201 \\
        (d) & 0.945 & 1.490 \\
        (e) & 0.789 & 2.652 \\
        (f) & 0.998 & 1.307
    \end{tabular}
    \caption{Best fit Luttinger liquid parameters for $\mathbb{Z}_3$ domain walls plotted in Fig.~\ref{fig.tnr-z3-coboundary}.}
    \label{tab.z3-coboundary-fit}
\end{table}

\subsection{Domain wall model of more complex Abelian group}
For gapless domain wall models with product of Abelian groups such as $G=\mathbb{Z}_2^2$, we find all of them have the same central charge $c=1$ and again can be described by Luttinger liquid theory with appropriate compactification radius $\rho$. This is not quite surprising because the gauge fluxes in these models all carry Abelian statistics, just like those simple $\mathbb{Z}_N$ models we studied above.

Below we consider a more interesting case with $G = \mathbb{Z}_2^3$. Under a proper choice of the 3-cocycle in this case, it turns out that the gauge flux will carry non-Abelian statistics. It is well known that for $G = \mathbb{Z}_N^3$, its third group cohomology $\mathcal{H}^3(\mathbb{Z}_N^3, U(1)) \cong \mathbb{Z}_N^7$ has seven generators.
Writing each group element as $g = (g^1, g^2, g^3)$ with $g^p = 0, \dots, N-1$,
the explicit expressions for the generators are\cite{hu2013twisted}
\begin{gather}
\alpha_{I}^{pq}(g_0, g_1, g_2, g_3) 
= \exp\left[i\frac{2\pi}{N^2} g^p_{10} \left(g^q_{21} + g^q_{32} - g^q_{31}\right)\right]
\nonumber
\\
\alpha_{II}(g_0, g_1, g_2, g_3) 
= \exp\left[i\frac{2\pi}{N} g^1_{10} \, g^2_{21} \, g^3_{32}\right]
\end{gather}
where
\begin{gather}
\allowdisplaybreaks[0]
1 \leq p \leq q \leq 3 , \nonumber \\
g^p_{ij} \equiv (g^p_i - g^p_j)~\mathrm{mod}~N ,  \\
g^p_i \in \{0,1,\dots,N-1\} .  \nonumber
\end{gather}
A 3-cocycle of $\mathbb{Z}_N^3$ can be written as
\begin{equation}
\nu_3 = k_{II} \alpha_{II} + \sum_{p, q} k_I^{pq} \alpha_I^{pq}
\end{equation}
for $k_I^{pq}, k_{II} = 0, \dots, N-1$.

\begin{figure}
    \includegraphics[width = 0.2 \textwidth]{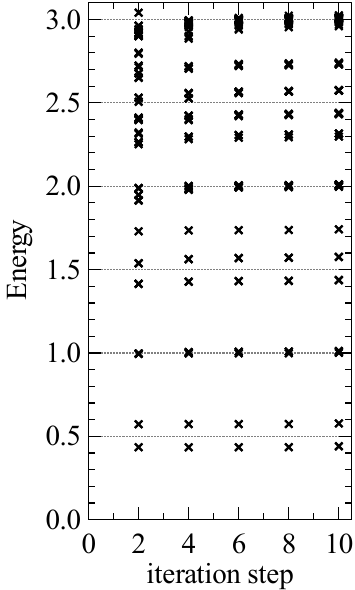}
    \caption{
        Scaling dimensions of a $\mathbb{Z}_2^3$ domain wall with a non-trivial $\alpha_{II}$ generator.
        Best fit compactification radius $\rho = 0.932$.
    }
    \label{fig.z222}
\end{figure}

We will restrict ourselves to the simplest case of $N=2$.
Of particular interest to us is the generator $\alpha_{II}$, which involves all three sub-$\mathbb{Z}_2$ groups.
We compute the scaling dimensions of the domain wall defined by the cocycle $k_I^{pq} = 0, k_{II} = 1$, which corresponds to the case with non-Abelian statistics for the gauge flux.
We again find such a gapless domain wall model can be described by a Luttinger liquid with $c=1$.
As seen in Fig.~\ref{fig.z222}, the data again agrees with a compactified boson CFT, with compactification radius $\rho=0.932$.

\subsection{Domain wall model of non-Abelian group}
Finally, we consider an example with non-Abelian gauge group $G$. It is well known that the smallest non-abelian group is $S_3$.
We label each group element by a pair of numbers $g = (g^1, g^2)$,
where $g^1 = 0, 1$ and $g^2 = 0, 1, 2$.
Group multiplication rule is
\begin{align}
g h = & (g^1, g^2) (h^1, h^2) \nonumber
\\
= & (\avg{g^1 + h^1}_2, \avg{(-1)^{h^1} g_2 + h_2}_3)
\end{align}
where $\avg{n}_m = n~\mathrm{mod}~m$.
The group cohomology $\mathcal{H}^3(S_3, U(1)) \cong \mathbb{Z}_6$.
An explicit formula of 3-cocycles is\cite{hu2013twisted}
\begin{align}
& \nu_3(g_0, g_1, g_2, g_3) \nonumber \\
= & \exp \Big\{ i \frac{2 \pi k}{9} \Big[a_{10} (a_{21} + a_{32} - a_{31}) 
+ \frac{9}{2} b_{10} b_{21} b_{32}
\Big]
\Big\}
\end{align}
where
\begin{equation}
\begin{gathered}
a_{ij} = (-1)^{g_i^1} \avg{g^2_i - (-1)^{g^1_i+g^1_j} g^2_j}_3
\\
b_{ij} = \avg{g_i^1 + g_j^1}_2
\end{gathered}
\end{equation}
and $k = 0, \dots, 5$ labels the six inequivalent 3-cocycles.

\begin{figure}
    \includegraphics[width=0.3 \textheight]{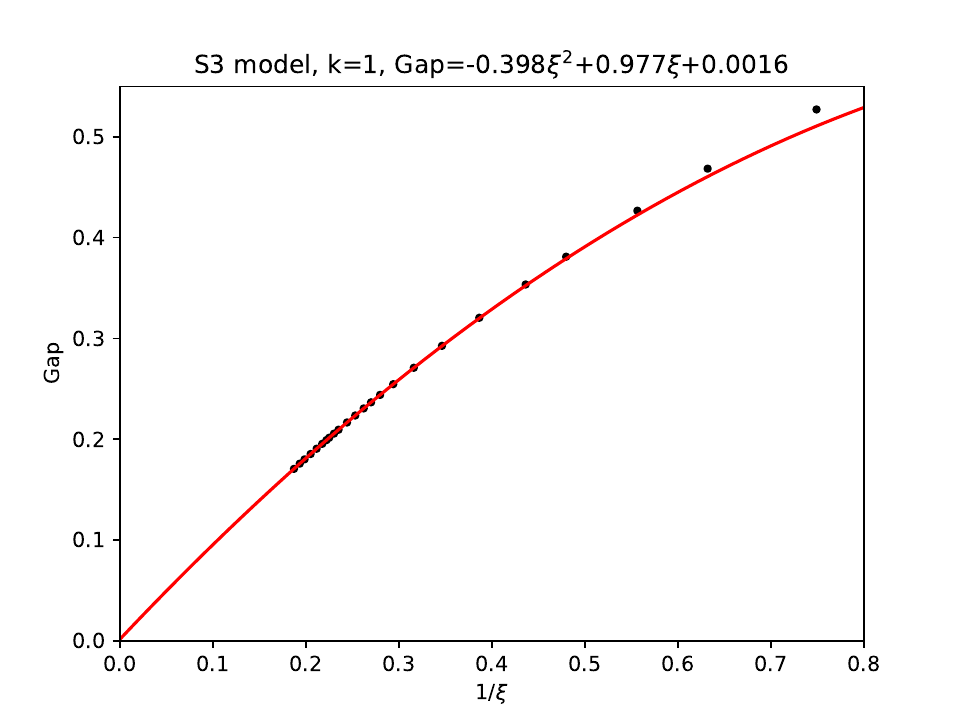}
    \caption{Finite size scaling of the $S_3$ gapless domain wall model.}
    \label{fig.gap-S3}
\end{figure}
\begin{figure}
    \includegraphics[width=0.3 \textheight]{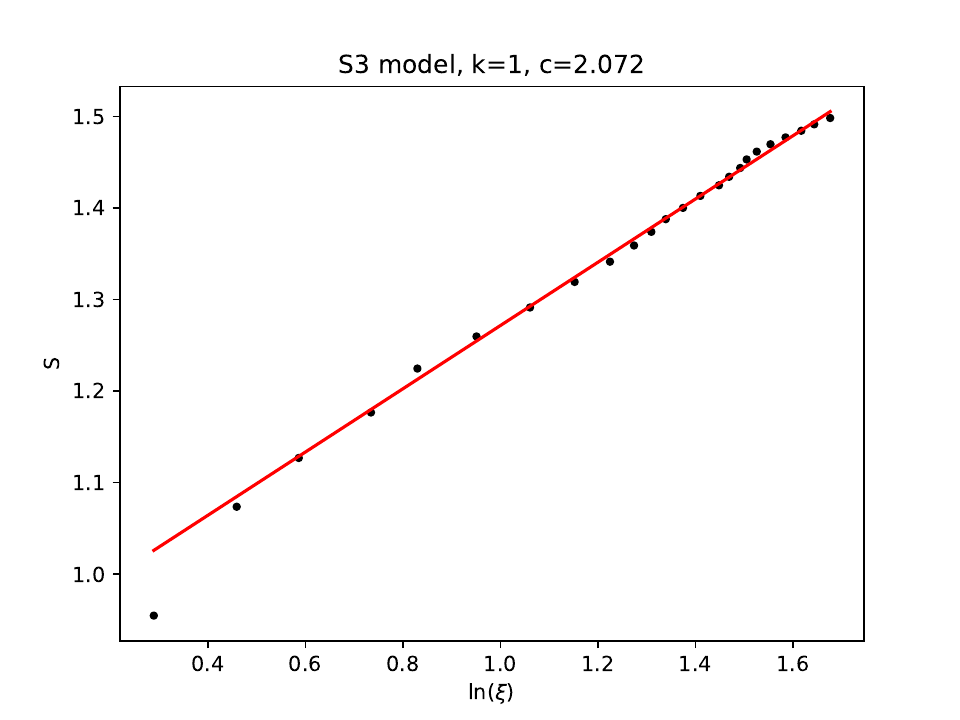}
    \caption{Entanglement entropy $S(\xi)$ of correlation length $\xi$ in the ground state of infinite size $S_3$ domain wall models with $k=1$. Data is fitted with $S(x) = c \mathrm{ln}(\xi)/6 + b$ and we find $c\sim 2.072$.}
    \label{fig.entropy-S3}
\end{figure}

We use the variational optimization algorithms for uniform matrix product states (MPS) \cite{zauner2018vumps,vanderstraeten2019vumps} to obtain the ground state. 
Given a translationally invariant MPS, the correlation length $\xi$ can be extracted from the spectral properties of the transfer operator \cite{verstraete2008review} $\mathbb{E}=\sum_{s} \bar{A}^{s} \otimes A^{s}$,
\begin{align}
\xi = -\frac{1}{\mathrm{ln}|\frac{\lambda_{2}}{\lambda_{1}}|}.
\end{align}
Here $\bar{A}^{s}$ is the complex conjugate of $A^{s}$. $\lambda_{1}$ and $\lambda_{2}$ are the largest and second largest eigenvalue of $\mathbb{E}$. The ratio between $\lambda_{1}$ and $\lambda_{2}$ also bounds the gap of the parent Hamiltonian \cite{schuch2013review,nachtergaele1996gap},
\begin{align}
\mathrm{Gap} &= 1- \frac{\lambda_{2}}{\lambda_{1}}.
\end{align}
The central charge $c$ of the CFT is related to correlation length $\xi$ by the scaling relation \cite{Calabrese2004entropy}
\begin{align}
S &= \frac{c}{6} \mathrm{ln} \xi
\end{align}
In MPS algorithms, both $S$ and $\xi$ are controlled by the numerical parameter ``bond dimension'' $D$. As $D$ increases, we obtain a more accurate ground state with larger entanglement entropy $S$ and larger correlation $\xi$. In Fig.~\ref{fig.entropy-S3}, we plot $S(\xi)$ with respect to $\xi$ for several $D$. From the data fitting we find $c\sim 2.072$. In Fig.~\ref{fig.gap-S3}, we plot the gap as a function of correlation length $\xi$. It shows the gap is closed as $\xi$ diverges, which confirms the gapless nature of the S3 model with $k=1$. 


Unfortunately, due to a very big truncation error, we can not implement the loop-TNR algorithm to compute the scaling dimension in this case. Possible candidate CFTs with $c=2$ include the $su(3)_1$ Wess-Zumino-Witten model and two-component Luttinger liquid theory with proper compactification radii.


\section{Conclusions and discussions}
In conclusion, we systematically construct lattice models of gapless domain walls between twisted and untwisted gauge models with arbitrary finite group $G$ in $2+1$D.
We then use the state-of-art loop-TNR algorithm to study several examples. For all the Abelain group cases studied here, we find all of them perfectly agree with the Luttinger liquid theory descriptions with a central charge $c=1$ at low energy. For the simplest non-Abelian group case with $G=S_3$, we find it is still gapless but has a larger central charge $c=2$. We conjecture that the $su(3)_1$ Wess-Zumino-Witten model could be a very good candidate for such a CFT.  
 
We further provide a physical picture to understand the gapless nature of these domain walls based on the theory of Lagrangian subsets and anyon condensation. On the other hand, since the corresponding ungauged domain walls can be regarded as an SPT boundary with an anomalous symmetry action, they must be gapless if the anomalous symmetry is not broken spontaneously, which is true for all domain wall models constructed in this paper. We also would like to stress that for a given finite group $G$, our construction actually gives rise to a deformation class of gapless domain wall models parameterized by different coboundary choices, and our numerical results have shown that all these models have the same central charge but with different compactification radius and velocity ratios in general.
  
Finally, it is straightforward to generalize our construction of gapless domain wall models into higher dimensions using the correspondence between SPT models and twisted gauge models.
All we need is a branched triangulation on a higher dimensional manifold, 
and an appropriate region $\mathcal{A}$ (Fig.~\ref{fig.general-dw}) to define the unitary transformation $U'$. 
It can be easily seen that for a higher dimensional domain wall Hamiltonian \( H = - \sum_i H_i \), 
each local term $H_i$ acts only on the site $i$ and its nearest neighbors.
But to write down an explicit model is rather tedious.
The difficulty partly comes from the \emph{reduced} translational and rotational symmetry when a branched triangulation is imposed on a higher dimensional manifold.
Such reduced symmetry also increases the difficulty for numerical study of these models. Physically, we believe that these models still describe gapless phases since the anomalous symmetry cannot be broken on domain walls separating two deconfinement phases of gauge group $G$. From a more mathematical point of view, all these domain wall models realize the so-called categorical symmetry. Thus, our construction might even give rise to a systematical way of understanding gapless domain walls in bosonic systems, especially in 2+1D, since recent studies have shown that all topological phases in 3+1D can be realized as twisted gauge theory\cite{Tian}. In future work, it will be of great interest to investigate the whether these gapless domain wall models are integrable or not. The generalization of our constructions into fermionic systems is also another important direction. 


\section*{Acknowledgments}
We thank A. W. W. Ludwig, D. N. Sheng and Davide Gaiotto for stimulating discussions and early collaborations on this project. We also thank M. Metlitski, J.-Q. Wu and Y.-H. Zhang for very helpful discussions on conformal field theories.
This work is supported by funding from Hong Kong’s Research Grants Council (GRF no.14301219) and Direct Grant no. 4053409 from The Chinese University of Hong Kong. S. Yang is supported by NSFC (Grant No. 11804181) and the National Key R\&D Program of China (Grant No. 2018YFA0306504). Work at Perimeter Institute is supported by the Government of Canada through the Department of Innovation, Science and Economic Development Canada and by the Province of Ontario through the Ministry of Research, Innovation and Science.

\appendix

\section{Algebraic definition of group cohomology}
\label{sec.cohomology}

For a group $G$, a $G$-module $M$ is itself an Abelian group 
on which the group $G$ acts compatibly with its Abelian group structure, i.e.
\begin{align}
g \cdot (ab) = (g \cdot a)(g \cdot b), \quad \forall g \in G, \quad \forall a, b \in M.
\end{align}

Define a $n$-cochain as a map $\nu_n: G^{n+1} \to M$ that satisfies\footnote
{It is common to define $n$-cochains as an arbitrary function from $G^n$ to $M$.
Our definition involves one more group element and an constraint on the function, 
and is essentially equivalent.
Such a definition, although with redundancy, 
allows more convenient definitions and intuitive geometric interpretations of 
many of the algebraic objects defined later.}
\begin{align}
g \cdot \nu_n(g_0, g_1, \dots, g_n) = \nu_n(g g_0, g g_1, \dots, g g_n).
\end{align}
The set of all $n$-cochains forms a group, denoted as $\mathcal{C}^n(G,M)$, 
whose group multiplication is simply the function multiplication of $\nu_n$.

Define the coboundary operator as a map 
\begin{align}
d_n : \, \mathcal{C}^n(G,M) \quad & \to \quad \mathcal{C}^{n+1}(G,M)
\nonumber \\
( d_n \nu_n ) ( g_0, \dots, g_n ) & = \prod_{i=0}^{n+1} \nu^{(-1)^i}(g_0, \dots, g_{i-1}, g_{i+1}, \dots, g_n)
\label{eq.cocyclecdn}
\end{align}
An $n$-cochain $\nu_n$ is called an $n$-coboundary if $ \nu_n = d_{n-1} \nu_{n-1}$ for some $\nu_{n-1} \in \mathcal{C}^{n-1}(G,M)$.
It is called an $n$-cocycle if $d_n \nu_n = 1 $.
The set of all $n$-coboundaries $\mathcal{B}(G,M)$ and 
the set of all $n$-cocycles $\mathcal{Z}_n(G,M)$ form two subgroups of $\mathcal{C}^n(G,M)$,
where we define $B^0(G,M) = 0 $ in addition.
More formally, we have
\begin{align}
\mathcal{B}^n(G,M) = & \begin{dcases}
1, &  n=0;  \\
d_{n-1}( \mathcal{C}^{n-1}(G,M)), & n \geq 1.
\end{dcases}
\\
\mathcal{Z}^n (G,M) = & \, \ker( d_n ).
\end{align}
Finally, we define the group cohomology of $(G,M)$ as the quotient group
\begin{align}
\mathcal{H}^n(G,M) = \mathcal{Z}^n(G,M)\, /\, \mathcal{B}^n(G,M).
\end{align}

For our purposes, it is sufficient to consider $M = U(1)$ whose elements are simply phase factors, 
and $G$ is the symmetry group of the system.
$G$ acts on $U(1)$ in the following way:
\begin{align}
g \cdot a = a^{s(g)}, \quad \forall g \in G, \quad \forall a \in M,
\end{align}
where
\begin{align}
s(g) = \begin{dcases}
1,   &  \text{if}\; g \in G \; \text{acts unitarily;}  \\
-1,  &  \text{if}\; g \in G \; \text{acts anti-unitarily.}
\end{dcases}
\end{align}
To explicitly indicate this nontrivial action of anti-unitary group elements, 
we shall from now on write $M= U_T(1)$.

\section{Constructing bosonic SPT phases using group cohomology}
Bosonic SPT phases can be systematically described by group cohomology theory \cite{bosonlong}.
Specifically, $(d+1)$-dimensional bosonic SPT phases with symmetry group $G$ 
can be labeled by elements in $\mathcal{H}^{d+1}[G, U_T(1)]$.
The identity element corresponds to a trivial phase (product state), 
and nontrivial elements correspond to nontrivial SPT phases.

\begin{figure}[h]
\centering
\includegraphics[width = 0.16\textwidth]{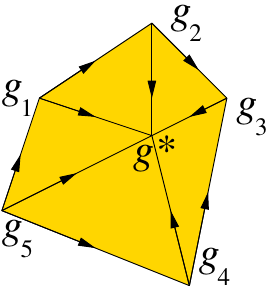}
\caption{
    The graphical representation of the ground state wave function \eqref{eq.Psi} in the case of $d=1$.
    The edge is $M$ and the disk is $M_{\mathrm{ext}}$.}
\label{fig.Psi}
\end{figure}

Furthermore, we can construct exactly solvable lattice models of SPT phases for any finite symmetry group $G$ and in any dimensions.
Such a model is constructed on a $d$-dimensional simplicial complex $M$,
which is itself the boundary of an extended $(d+1)$-dimensional complex $M_{\mathrm{ext}}$ with a branching structure,
and we further assume that there is only one internal vertex in $M_{\mathrm{ext}}$.
Each vertex in $M$ is associated with a $\abs{G}$-dimensional local Hilbert space, 
where basis vectors $\ket{g_i}$ are labeled by group elements $g_i \in G$.
The internal vertex of $M_{\mathrm{ext}}$ is associated with a fixed group element $g^* \in G$.
The ground state wave function is given by (Fig.~\ref{fig.Psi})
\begin{align}   \label{eq.Psi}
\largeket{\Psi_M} = \sum_{\{g_i\}_M} \prod_{\{ij \dots *\}} 
        \nu_{d+1}^{s_{ij \dots *}} (g_i, g_j, \dots , g^*) \largeket{ \{g_i\}_M }
\end{align}
where the sum runs over all configurations $\{g_i\}_M$ of the vertices in $M$ and 
the product runs over all simplices $\{ij \dots *\}$ in $M_{\mathrm{ext}}$,
and $\nu_{d+1}$ is a group $(d+1)$-cocycle.
The symmetry is on-site and acts in the following simple way:
\begin{align}
g: \largeket{ \{g_i\}_M } \to  \largeket{ \{g g_i\}_M }, \quad g \in G
\end{align}

The ground state is trivial a product state when the cocycle is trivial.
Nontrivial ground states can be obtained from this trivial state by a unitary transformation
\begin{align}  \label{eq.unitary}
U = \prod_{\{ij \dots *\}} \nu_{d+1}^{s_{ij \dots *}} (g_i, g_j, \dots , g^*)
\end{align}
An exactly solvable Hamiltonian can be constructed for this ground state as 
\begin{align}  \label{eq.Hi}
H = - \sum_i H_i, \quad H_i = U \outerp{\phi_i}{\phi_i} U^\dag
\end{align}
where the sum runs over all vertices $i$ and $\ket{\phi_i} = \sum_{g_i \in G} \ket{g_i}$.
It is straightforward to check that all $H_i$'s commute with each other, 
hence the solvability of this model.

\begin{figure}[bt]
\centering
\includegraphics[width = 0.4\textwidth]{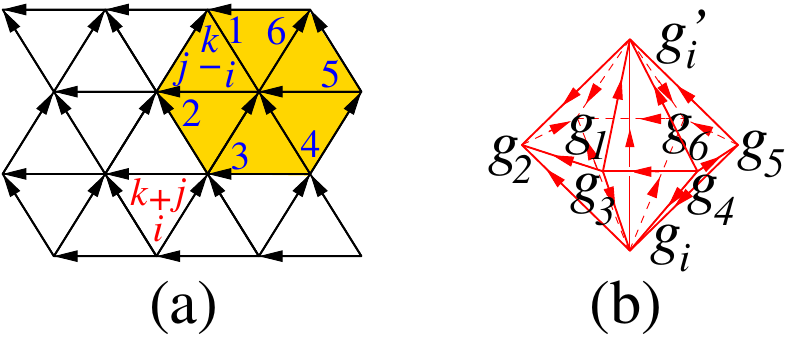}
\caption{
    A lattice model for 2-dimensional SPT phase.
    (a) $H_i$ acts on seven spins centered around $i$, shown as the shaded region.
    Down triangles $\triangledown$ have orientation $s_{ijk}=+$, 
    and up triangles $\triangle$ have orientation $s_{ijk}=-$.
    (b) A graphical representation of the phase factor, which is a product of $3$-cocycles.
    The ``internal'' vertex associated with $g^*$ was originally in the center of the cage, 
    but can be removed using cocycle condition (either graphically or algebraically).%
}
\label{fig.2d-general}
\end{figure}

In 2D, an explicit formula is given by (Fig.~\ref{fig.2d-general}):
\begin{align}
& \matrixel{g_i', g_1 g_2 g_3 g_4 g_5 g_6}{H_i}{g_i, g_1 g_2 g_3 g_4 g_5 g_6} \nonumber \\ 
= & \frac{ \nu_3(g_4,g_5,g_i,g_i') \nu_3(g_5,g_i,g_i',g_6) \nu_3(g_i,g_i',g_6,g_1) }
    { \nu_3(g_i,g_i',g_2,g_1) \nu_3(g_3,g_i,g_i',g_2) \nu_3(g_4,g_3,g_i,g_i') }
\end{align}
For $G=\mathbb{Z}_2$, there are two elements in $\mathcal{H}^3[\mathbb{Z}_2,U_T(1)]$, 
corresponding to the two Hamiltonians defined by \eqref{eq.H}.

Finally, we note that elements in the group cohomology are equivalence classes.
Two different cocycles $\nu_n$ and $\nu_n'$ describe the same SPT phase if they differ only by a coboundary:
\begin{align}
 & \nu_n (g_0,g_1, \dots, g_n) \nonumber \\
= & ( d_{n-1} \nu_{n-1} ) (g_0,g_1, \dots, g_n) \cdot \nu_n' (g_0,g_1, \dots, g_n)
\end{align}
The choice of the branching structure is also irrelevant in the classification of SPT phases,
so we can always choose any branching structure that is most convenient for our purpose.

\section{Energy spectrum of Luttinger liquid theory}
\label{sec:luttinger}

In this appendix, we derive the energy spectrum \eqref{eq.chiral-boson-energies} for the $c=1$ Luttinger liquid theory whose action is given in Eq.~\eqref{eq.LL}. For convenience, we repeat the action here
\begin{equation}
S_{\mathrm{edge}} = \frac{1}{4\pi} \int dt dx \left( K_{IJ} \partial_t \phi_I \partial_x \phi_J - V_{IJ} \partial_x \phi_I \partial_x \phi_J \right)
\label{act-app}
\end{equation}
where $\phi_I$ is a compact boson with $\phi_I=\phi_I+2\pi$. As discussed in the main text, we only need to consider the $c=1$ two-component theory associated with
\begin{equation}
K=\left(
\begin{matrix}
0 & 1 \\
1 & 0
\end{matrix}
\right), \quad  V=\left(
\begin{matrix}
a & c \\
c & b
\end{matrix}
\right)
\end{equation}
where $a,b,c$ are real numbers satisfying the conditions $a>0$ and $ab>c^2$ such that $V$ is  positive definite. This is a free theory and it is well studied in the literature (see e.g. Ref.~\onlinecite{CFT-book}). To be self-contained, we give a brief derivation on the relevant results used in the main text, with a focus on the case that the left and right movers have \emph{different} velocities.

To find the spectrum, we perform the following change of variables:
\begin{equation}
\label{change-var}
\tilde\phi=U^{-1}{\phi}
\end{equation}
where $\phi=(\phi_1,\phi_2)$ and
\begin{equation}
U = \frac{1}{\sqrt{2}}\left(\begin{matrix}
\rho & -\rho\\[5pt]
1/\rho & 1/\rho
\end{matrix}
\right)
\label{U-mat}
\end{equation}
with $\rho=\sqrt[4]{b/a}$. Inserting \eqref{change-var} into \eqref{act-app}, the action is rewritten as
\begin{align}
S_{\mathrm{edge}} = & \frac{1}{4\pi} \int dtdx \ \partial_x\tilde{\phi}_1\left( \partial_t\tilde\phi_1 -v_1\partial_x\tilde\phi_1 \right) \nonumber \\ 
& - \frac{1}{4\pi} \int dtdx \  \partial_x\tilde{\phi}_2\left(\partial_t\tilde\phi_2 +v_2\partial_x\tilde\phi_2 \right)
\label{act-app2}
\end{align}
where $v_1= \sqrt{ab}+c$ and $v_2=\sqrt{ab}-c$. The two fields $\tilde{\phi}_1$ and $\tilde{\phi}_2$ completely decouple, each of which is a standard free chiral boson. The theory \eqref{act-app2} can then be solved using the standard mode expansion; see e.g. Ref.\onlinecite{CFT-book}. Below we briefly state the main results. 

The energy eigenstates consist of two types. First, there is a set of highest wight states, created by the corresponding vortex operators when acting on the ground state. This energy of these states depend on the compactification radii of the fields $\tilde\phi_1$ and $\tilde\phi_2$, which inherit from those of $\phi_1$ and $\phi_2$. In the current case that $\phi_I=\phi_I + 2\pi$, the general form of vortex operators are given by
\begin{equation}
V_{m,n} = e^{im\phi_1 + i n\phi_2}
\end{equation}
where $m,n$ are integers. Acting $V_{m,n}$ on the ground state $|0\rangle$, it creates a highest weight state $|m,n\rangle \equiv V_{m,n}|0\rangle$. These states are the primary states of the $U(1)$ Kac-Moody algebra that one can read off from the action $S_{\rm edge}$. We rewrite $V_{m,n} = e^{i\tilde{m}\tilde\phi_1 + i\tilde{n}\tilde{\phi}_2}$, 
where
\begin{equation}
\tilde{m} = \frac{1}{\sqrt{2}}(m \rho + n/\rho), \quad  \tilde{n} = \frac{1}{\sqrt{2}}(-m \rho + n/\rho)
\end{equation}
The energy of the state $|m,n\rangle $ (relative to the ground state) is given by
\begin{align}
\Delta_{m,n} & = \frac{\pi v_1}{L}\tilde{m}^2 + \frac{\pi v_2}{L}\tilde{n}^2 \nonumber\\
&  = \frac{\pi(v_1+v_2)}{L}\left[\frac{1}{2}\left(\rho^2 m^2 + \frac{n^2}{\rho^2}\right) + \frac{1-v_r}{1+v_r}mn \right]
\end{align}
where $L$ is the system size and $v_r= v_2/v_1$.

Second, in the mode expansion of the fields $\tilde{\phi}_1$ and $\tilde\phi_2$, one introduces the Fourier coefficients $\tilde a_{1,l}$ and $\tilde a_{2,l}$, where $l$ is integer. The highest weight states are annihilated by $\tilde a_{1,l}$ and $\tilde a_{2,l}$ with $l>0$, i.e., $\tilde a_{1,l}|m,n\rangle = \tilde a_{2,l}|m,n\rangle =0$. On the other hand, a Fork space is spanned by acting $\tilde a_{1,l}$ and $\tilde a_{2,l}$ with $l<0$ on each primary state $|m,n\rangle$:
\begin{equation}
\tilde a_{1,-1}^{n_{1,1}} \tilde a_{1,-2}^{n_{1,2}}\dots  \tilde a_{2,-1}^{n_{2,1}} \tilde a_{2,-2}^{n_{2,2}}\dots |m,n\rangle
\end{equation}
where $n_{1,l}$ and $n_{2,l}$ are positive integers. The energy of these descendant states are
\begin{align}
\Delta_{m,n,\{n_{1,l},n_{2,l}\}} & = \Delta_{m,n} + \sum_{l=1}^{\infty}\left( \frac{2\pi v_1 l}{L} n_{1,l}+ \frac{2\pi v_2 l}{L} n_{2,l}\right) \nonumber\\
&  = \Delta_{m,n} + \frac{\pi (v_1+v_2)}{L} \frac{2}{1+v_r}\sum_{l=1}^\infty l(n_{1,l}+v_rn_{2,l})
\end{align}
Then, the whole spectrum is generated by varying the integers $m,n$, and $\{n_{1,l}, n_{2,l}\}$.

Finally, we comment that the ground state energy also depends on the system size $L$. For periodic boundary conditions, the ground state energy is given by
\begin{equation}
E_0 = -\frac{\pi c}{12 L}(v_1+v_2) 
\label{gs}
\end{equation}
where $c$ is the central charge. Since $c=1$, Eq.~\eqref{gs} can be used to set the energy unit $\pi(v_1+v_2)/L$ in numerical calculations.

\section{Virtual-space transfer matrix of quantum models}
\label{sec.comp-details}
In this section, we briefly explain how virtual energies are defined and computed using loop-TNR.

\begin{figure}
    \includegraphics[width=0.48\textwidth]{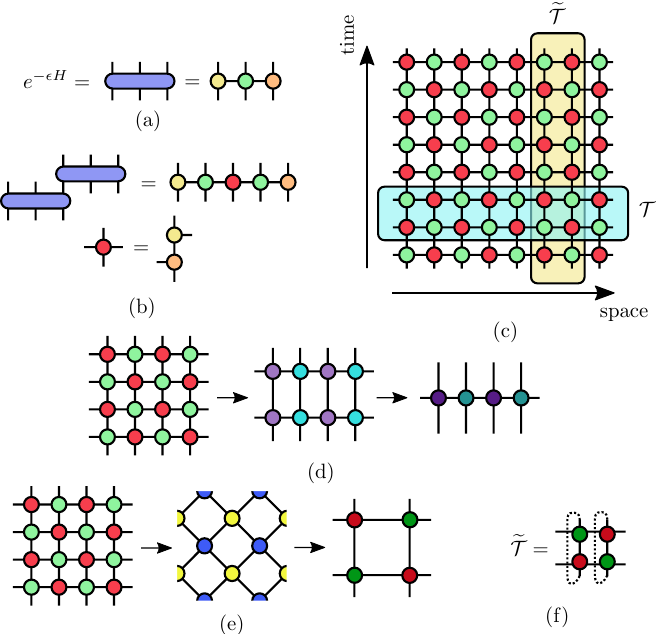}
    \caption{
        (a) Local evolution operator represented as a rank-6 tensor, and decomposed into an MPO via singular value decomposition.
        (b) Within a single layer of even or odd evolution operators, local evolution operators are connected to form a larger MPO.
        (c) A square network of rank-4 tensors representing Euclidean path integral of a 1D quantum system. A real-space transfer matrix $\mathcal{T}$ and a virtual-space transfer matrix $\widetilde{\mathcal{T}}$ are marked by shaded regions.
        (d) Compress the tensor network in the temporal direction by iTEBD.
        (e) Two iterations of loop-TNR iterations. Each tensor in the third diagram is a coarse-grained representation of four tensors in the first diagram.
        (f) A virtual-space transfer matrix constructed from four coarse-grained local tensors.
    }
    \label{fig.TN}
\end{figure}

The Euclidean path integral of a 1D quantum lattice system can be represented as a 2D tensor network through Suzuki-Trotter expansion\cite{suzuki1976trotter}. For our domain wall models, Hamiltonians, e.g. Eq.~\eqref{eq.H-dw-spec-branch}, consist of three body interaction terms, but local terms do not commute $[H_m, H_n] \neq 0$ only when $n=m\pm 1$. We therefore separate local terms into groups of even and odd terms
\begin{gather}
H = H_{\mathrm{e}} + H_{\mathrm{o}} \nonumber \\
H_{\mathrm{e}} = \sum_{n \text{~even}} H_n , \qquad H_{\mathrm{o}} = \sum_{n \text{~odd}} H_n
\end{gather}
All $H_n$'s commute within each group. A small local evolution can be represented as a rank-6 tensor
\begin{align}
T_{s_{n-1}s_ns_{n+1}}^{s'_{n-1}s'_ns'_{n+1}} = \matrixel{s'_{n-1}s'_ns'_{n+1}}{e^{-\epsilon H_n}}{s_{n-1}s_ns_{n+1}}.
\end{align}
The Euclidean time evolution operator is well approximated by
\begin{align}
U(\beta) \approx \left[ e^{-\epsilon H_{\mathrm{e}}} e^{-\epsilon H_{\mathrm{o}}} \right]^{\beta/\epsilon}, \quad \epsilon \ll 1
\end{align}
which is a stack of $2\beta/\epsilon$ alternating layers of even/odd local evolutions. Each layer can be written in the form of matrix product operators (MPO). The end result is a square network of rank-4 tensors. See Fig.~\ref{fig.TN}a-c.

The partition function $Z = \Tr~e^{-\beta H}$ of a periodic 1D quantum system is obtained by putting the tensor network in Fig.~\ref{fig.TN}c on a torus.
Following Ref.~\onlinecite{suzuki1985transfer}, we define the real-space transfer matrix $\mathcal{T}$ as a time evolution operator, i.e. a strip of the tensor network that wraps around in the spatial direction and transfer in the temporal direction. Similarly, we define the virtual-space transfer matrix $\tilde{\mathcal{T}}$ as a strip of the tensor network that wraps around in the temporal direction and transfer in the spatial direction (Fig.~\ref{fig.TN}c).

The virtual-space transfer matrix effectively defines a related quantum system on the virtual/Trotter space, where it serves as the evolution operator. \emph{Virtual energies}, as the name suggests, are energies of this related virtual space quantum system, which are real exponents $\tilde{E}_j$ of eigenvalues of the virtual-space transfer matrix $\tilde{\lambda}_j = e^{-l (\tilde{E}_j+i\tilde{P}_j)}$, where $l$ is a normalization constant. The imaginary exponent $P_j$ does not necessarily have physical meaning, and can be eliminated by taking appropriate powers of $\tilde{\mathcal{T}}$.

Relation between the original quantum system and the virtual system can be understood via their effective field theories, which share the same partition function but have switched roles of spatial and temporal directions. In other words, they are related via an $\mathcal{S}$ modular transformation.
A CFT realized on a lattice is modular invariant, so we expect virtual energies to provide the same information as energies of the original system. For non-conformal theories, the effect of $\mathcal{S}$ transformation can also be analyzed relatively easily.

An important distinction between the original lattice system and its related virtual system is that the former may have \emph{non-analytic} finite size corrections due to the discreteness of the lattice\cite{woynarovich1989nonanalytic}. Such corrections may render many system sizes unsuitable for taking the continuum limit, severely obscuring the corresponding field theory. On the other hand, the virtual system is defined on a continuous strip with no inherent discrete structure, and does not suffer from this effect. This is why we compute virtual energies in our study of domain walls.

To efficiently compute virtual energies, we first use iTEBD\cite{orus2008iTEBD} to compress the network in the temporal direction (Fig.~\ref{fig.TN}d) so that each local tensor becomes less anisotropic and spans a time interval of order 1. Then we use loop-TNR\cite{loop-TNR} to iteratively coarse-grain the square tensor network, so that a single local tensor covers exponentially larger area of Euclidean space-time (Fig.~\ref{fig.TN}e). After 10-20 iterations, the virtual-space transfer matrix constructed from just a few local tensors (Fig.~\ref{fig.TN}f) is enough to give results close to the thermodynamic limit. Virtual energies are found by sparse diagonalization of the virtual-space transfer matrix.

\bibliography{references}

\end{document}